\documentclass{aa}

\usepackage{longtable}
\usepackage{graphicx}
\usepackage{subcaption}
\usepackage{float}
\usepackage{color}
\definecolor{deepblue}{rgb}{0,0,0.5}
\definecolor{deepred}{rgb}{0.6,0,0}
\definecolor{deepgreen}{rgb}{0,0.5,0}

\usepackage{listings}
\newcommand\pythonstyle{\lstset{
language=Python,
basicstyle=\ttm,
morekeywords={self},              
keywordstyle=\ttb\color{deepblue},
emph={MyClass,__init__},          
emphstyle=\ttb\color{deepred},    
stringstyle=\color{deepgreen},
frame=tb,                         
showstringspaces=false
}}

\lstnewenvironment{python}[1][]
{
\pythonstyle
\lstset{#1}
}
{}

\newcommand\pythoninline[1]{{\pythonstyle\lstinline!#1!}}

\graphicspath{{./}{figures/}}

\usepackage[varg]{txfonts}
\begin{document}

\title{Accelerations of stars in the central 2-7 arcsec from Sgr~A*}
\institute{European Space Agency, European Space Astronomy Center, Madrid, Spain\label{esac}
\and 
Max-Planck-Institute for Extraterrestrial Physics,Gie{\ss}enbachstr. 1, D-85748 Garching, Germany\label{mpe}
\and
Sterrewacht Leiden, Leiden University, Postbus 9513, 2300 RA Leiden, The Netherlands\label{leiden}
\and
Departments of Physics and Astronomy, Le Conte Hall, University of California, Berkeley, CA 94720, USA\label{berkley}
\and
Max-Planck-Institute for Astrophysics, Karl-Schwarzschild-Straße 1, D-85748 Garching, Germany\label{mpa}
\and 
Excellence Cluster ORIGINS, Boltzmannstr. 2, 85748 Garching, Germany\label{origins}
\and
Max-Planck-Institute for Radio Astronomy, auf dem H{\"u}gel 69, D-53121 Bonn, Germany\label{mpir}
}
\author{A.~Young\inst{\ref{esac},\ref{mpe}}\and S.~Gillessen\inst{\ref{mpe}}\and
T.~de Zeeuw\inst{\ref{leiden},\ref{mpe}}\and Y.~Dallilar\inst{\ref{mpe}}\and A.~Drescher\inst{\ref{mpe}}\and F.~Eisenhauer\inst{\ref{mpe}}\and R. ~Genzel\inst{\ref{mpe},\ref{berkley}}\and F.~Mang\inst{\ref{mpe}}
\and T.~Ott\inst{\ref{mpe}}\and J.~Stadler\inst{\ref{mpa},\ref{origins}}\and O.~Straub\inst{\ref{origins},\ref{mpe}}\and S. ~von Fellenburg\inst{\ref{mpir},\ref{mpe}} \and F.~Widmann\inst{\ref{mpe}}}

\date{Received 13 September 2022 /
Accepted 25 November 2022}



\abstract{This work presents the results from extending the long-term monitoring program of stellar motions within the Galactic Center to include stars with separations of 2 - 7 arcseconds from the compact radio source, Sgr~A*. In comparison to the well studied inner 2 arcsec, a longer baseline in time is required to study these stars. With 17 years of data, a sufficient number of positions along the orbits of these outer stars can now be measured. This was achieved by designing a source finder to track the positions of $\sim 2000$ stars in NACO/VLT adaptive-optics-assisted images of the Galactic Center from 2002 to 2019. Of the studied stars, 54 exhibit significant accelerations toward Sgr~A*, most of which have separations of between 2 and 3 arcseconds from the black hole. A further 20 of these stars have measurable radial velocities from SINFONI/VLT stellar spectra, which allows for the calculation of the orbital elements for these stars, thus increasing the number of known orbits in the Galactic Center by $\sim40\,\%$. With orbits, we can consider which structural features within the Galactic Center nuclear star cluster these stars belong to. Most of the stars have orbital solutions that are consistent with the known clockwise rotating disk feature. Further, by employing Monte Carlo sampling for stars without radial velocity measurements, we show that many stars have a subset of possible orbits that are consistent with one of the known disk features within the Galactic Center. 
}
\maketitle
\titlerunning{Accelerations of stars in central 2-7 arcsec from Sgr~A*}
\authorrunning{A.~Young et al.}
\section{Introduction} \label{sec:intro}
For over a quarter century, near-infrared (NIR) astronomy has been successfully used to observe the nearest galaxy center available to astronomers, that of the Milky Way. The Galactic Center (GC) was discovered to be an incredibly rich stellar environment \citep{genzel1994, genzel1997,genzel1996darkmass, ekadngenzel1996, ghez1998, Ghez_2003, schodel_2002_firstorbit}, which allows it to serve as a laboratory for studying the fundamental astrophysical properties of environments governed by the potential of a central supermassive black hole (SMBH). One ongoing experiment involves obtaining measurements for the positions of the stars within the inner 2 arcseconds surrounding Sgr~A* using a combination of adaptive optics (AO) and Speckle data sets spanning from 1992 to 2019 paired with the enhanced resolution of the interferometric data collected using the GRAVITY instrument on the Very Large Telescope Interferometer (VLTI) from 2017 to 2022 \citep{schodel_2002_firstorbit,stellardynSchodel2003,Ghez_2003,Ghez_2008,Eisenhauer_2005,Gillessen2009,Gillessen2017,gravityredshift2018,GRAVITYmassdist2021}. In total, orbital determinations have been performed for $\sim 50$ stars in the GC, all of which contribute to the understanding of the GC environment \citep{schodel_2002_firstorbit,stellardynSchodel2003,Ghez_2003,Ghez_2008,Eisenhauer_2005,Gillessen2009,Gillessen2017,Meyer2012,Boehle2016}. This work continues the monitoring study of stars in the GC by expanding the projected outer search radius to the 7 arcseconds surrounding Sgr A*, a regime occupied by the ``clockwise disk'' \citep{2disksPaumard2006,Bartko2010,Yelda2014,Sebyoungstars}.

In particular, we utilize data collected from 2002 to 2019 using the combined Nasmyth Adaptive Optics System (NAOS)- Near-Infrared Imager and Spectrograph (CONICA) instrument better known as NACO and the Spectrograph for INtegral Field Observations in the Near Infrared (SINFONI) mounted on the European Southern Observatory's Very Large Telescope (ESO VLT) to track stellar motions for $\sim 2000$ stars. Sect. \ref{sec:data} provides an overview of the NACO AO data utilized to detect accelerations as well as the SINFONI datacubes that contain the stellar spectra. Sect. \ref{sec:sourcefinder} details the source finding algorithm developed to track the motions of the stars within the field. Sect. \ref{sec:accelerations} discusses the methods employed to detect the stellar accelerations. Sect. \ref{sec:orbitalelements} describes two methods for deriving orbital motions for the stars studied. The first pertains to stars that have measurable radial velocities from the SINFONI datacubes, while the second introduces a method that uses Monte Carlo sampling over a flat radial velocity prior for stars that do not have measured radial velocities. Sect. \ref{sec:gcenvironment} relates the results from this study to previous works that focused on the structure of the nuclear star cluster environment represented by the GC. Finally, a summary of results as well as suggestions for future observations that would further this study are provided in Sect. \ref{sec:summary}. 
\section{Data}\label{sec:data}
This work employs data from two instruments, namely NACO and SINFONI, that have since been decommissioned from ESO VLT in Paranal, Chile. The NACO images used in this work utilize the 13.3\,mas pixel scale. It is typical for a data set to contain approximately 2 hours of data, with a single-detector integration time of $\sim 15\,s$ \citep{Gillessen2009}. After integrating a few times on one field of view (FOV) it would be changed to a new pointing. There were typically four pointings used in total, such that the central 4\,arcseconds appear in all frames \citep{Gillessen2009}. The reduction of the data had already been performed  (see, e.g., \citealt{schodel_2002_firstorbit, warpedBartko2009, Gillessen2009, Gillessen2017}); however, the process generally followed, which includes sky subtraction and flat fielding, is outlined in \cite{Gillessen2009}. The reduction also includes processes that separate sources, namely moderate deconvolution in the central 5 arcseconds with the Lucy-Richardson algorithm \citep{Gillessen2009}. In total, 85 epochs were selected from the archive of NACO data based on a visual inspection looking for the best images from each year between 2002 to 2019, a list of which is included in Tables \ref{tab:NACOdata1} and \ref{tab:NACOdata2}.

Also key to this work, SINFONI is a NIR (1.1- 2.45\,micron) spectrograph that operated from 2004 until 2019  \citep{Eisenhauer2003sinfoni, SINFONImanual}. SINFONI is composed of two instruments, namely MACAO, which is an AO module, and the Spectrometer for Infrared Faint Field Imaging \citep[SPIFFI;][]{Eisenhauer2003sinfoni,bonnet_macao2003}. Again, data cubes utilized in this work have been previously analyzed (e.g., \citealt{Eisenhauer_2005, 2disksPaumard2006, warpedBartko2009, sfhgcPfuhl2011}). The majority of measurements were taken from a combined cube featuring observations from 18-19 August 2004. In this case, the observations have undergone a correction to remove features from the Earth's atmosphere. Additionally, this cube is useful because it has a FOV of 8\,arcsec $\times$ 8\,arcsec, meaning most of the stars studied in this work are contained in this single cube. For stars that were not included in the FOV covered by this cube, or if the spectra are too noisy, a tool developed by \cite{Sebyoungstars} was used to identify cubes with smaller FOVs featuring the relevant source. A table containing all of the stacked exposures is provided in Table \ref{tab:SINFONIdata}. 
\section{The source finder}\label{sec:sourcefinder}
In order to create a source finder, we selected a base image that we used to define the list of sources that were tracked over the 85 NACO images. The image 20110330 was selected due to the fact that it is the best image closest to the mean epoch of the sample, which will reduce correlations in the free parameters later on. We also considered how to best mitigate the deconvolution effects that occur around bright sources and can be misidentified as sources by the finder. We chose to perform image multiplication using three of the best NACO images including the base epoch and one from each of the 2010 and 2012 observing runs. By multiplying these images together we minimize the number of ``fake'' sources incorrectly identified by the finder. The images multiplied together were selected because they are the closest in time to the base epoch, thus minimizing the movement of the sources between the multiplied epochs. After image multiplication was performed, we used the Python package \texttt{DAOStarFinder} \citep{DAOstarfinder} to compile a list of sources to search for in all of the epochs. In total we detect 1839 sources in the multiplied image outside the central 2 arcseconds. The next step toward creating a source finder algorithm that is successful for multiple epochs is to establish an image stacking procedure. If the images are stacked on top of each other, accounting for slightly different pointings and telescope rotation, the pixel search grid can be limited to less than 2 pixels. This is a generous limit, since stars with separations greater than 2 arcseconds move significantly slower than the innermost stars. At this separation, stars generally have proper motions corresponding to less than one pixel between the epochs studied.

To stack the images and account for offsets in pointings and telescope rotations between epochs, we utilized third-order polynomial transformations to translate between the different NACO images and project the previously determined location for each source into the new epoch. As such, each epoch requires its own unique set of polynomial coefficients to perform this transformation. In order to calculate these coefficients, we identified a set of stars that appear in all epochs. In this case, 91 reference stars are used, which are the same as those defined in \cite{Gillessen2009} and \cite{irlocPlewa2015}. The position of each reference star is determined by hand for the base image then searched for specifically at all other epochs using a simplified version of the finder. Since these are particularly bright stars in the field, the finder performs very well which meant we could use simpler stacking techniques that do not require the set of transformations being determined. In this case, simply ensuring the positions of two bright sources are shifted such that they stack on top of the locations determined for the sources in the previous epoch was sufficient to obtain the remaining reference star locations.

Once these pixel positions were known for each epoch, we could solve for polynomial coefficients that transform the pixel positions from one observation to another. In this case, two sets of third-order polynomials were generated for each epoch. The first set advances the positions by one epoch, while the other advances the positions by the number of epochs since the 20110330 epoch. This is done in order to minimize the number of ``lost'' stars between epochs, or the number of stars for which no position can be determined by the finder in the new epoch. Essentially, if the star was detected in the previous epoch, the transformation coefficients from that epoch were used to project the pixel position to the pixel coordinates of the following epoch. These then served as an initial guess for the finder to determine the stellar position in the new epoch. If the star was not detected in the previous epoch, however, then the position of the star in the 20110330 image was transformed with the second set of coefficients to the new pixel coordinate system and again served as the initial guess for the finder. Stars can be lost in interim epochs either due to being a faint source in poor observing conditions or due to a confusion event. It is not possible to simply relate all epochs only using the polynomials that transform from the 20110330 image, as this set of coefficients is less successful at providing a good initial guess for the finder as the time between epochs increases. This is due to the motions of the sources that cannot be accounted for a priori \footnote{It would be possible to account for this motion by developing an iterative method that determines adjusted positions and proper motions for each star after every epoch. However, due to the small motion of stars at these projected separations and the fact that very few stars are lost between epochs, this was deemed unnecessary.}. As such, fewer stars from the original list of 1839 are lost for the epochs closest to 20110330 as this secondary set of transformations is often successful at finding the star if it is detectable. Within the nine epochs selected from 2011, on average the loss of stars between each epoch is 0.6\,\%. Whereas, within the six epochs used from 2002 - 2003, the average loss of stars between each epoch is approximately 2\,\%. This increase is due to the fact that as the number of years increases from this mean epoch, the transformation often does not provide a sufficiently good guess to locate lost stars between epochs due to their stellar motion over time. This highlights another reason to select the mean epoch of the sample as the base image. We also note that although stars that appear to be moving fast may be preferentially removed due to the design of the source finder, this will not bias the sample of detected physical accelerations. Since at these separations the average motion between epochs should be less than a pixel, the search box utilized by the source finder is sufficient to detect stars moving with physical motions. In this case, losses of stars with physical motions that occur are mostly because they are faint stars lacking strong contrast with the background of the NACO observation. Common losses also include the loss of the stars located on the edge of the FOV that may be lost between epochs due to slightly different pointings. 

\begin{figure}[h]
    \centering
    \includegraphics[width=.45\textwidth]{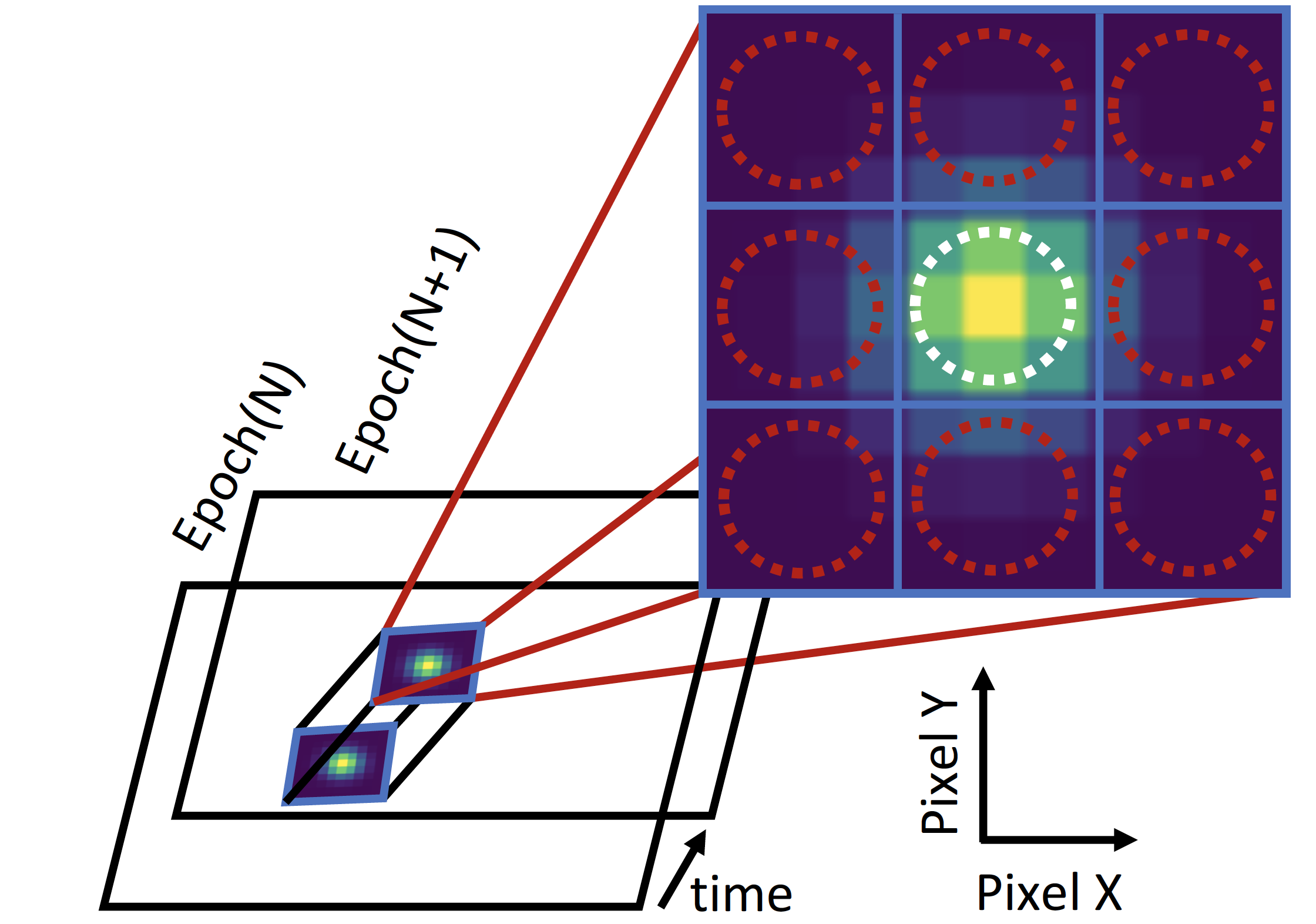}
    \caption{Illustration of the source finding algorithm. First, the position from Epoch(N) is projected to Epoch(N+1) using the polynomial transformations. The transformed position is then used as an initial guess for the finder. Each grid square in the zoomed-in panel, which shows a NACO source profile, corresponds to a Gaussian search box of some side length. The fitter tries fitting a Gaussian profile to the pixels in each of the surrounding grid squares. The Gaussian fit with the lowest $\chi^2$ (\textit{white circle}) is then selected as the position for this star in the new epoch. For clarity, the search boxes appear here as a grid; however, the actual algorithm utilizes overlapping search boxes.}
    \label{fig:gausfindergrid}
\end{figure}

Once there was an initial guess for the pixel position of the star, this was fed into the finder. The finder requires the definition of a fitting box to perform a Gaussian fit to the data contained in the box. To account for the fact that the sources do not appear uniform in size due to photon spilling in the CCD detector, we used multiple side lengths for this fitting box in iteration. These side lengths typically ranged from 8 to 16 pixels to get a good fit. Additionally, the finder performs a fit at each pixel moving outward from the initial guess position by 0 to 2 pixels in every direction. A simplified schematic of the process is shown in Fig. \ref{fig:gausfindergrid}.

The fit was performed by \texttt{LevMarLSQFitter}, which utilizes the Levenberg-Marquardt algorithm and least squares statistic to return a best-fit result \citep{LEVMarq}. The fit corresponding to the lowest $\chi^2$ was then selected. We verified the new position of the stars by visual inspection of the resulting pixel positions plotted onto the NACO image of the new epoch. We find that the source finder is capable of consistently determining the location of the source in a new epoch as required.
\section{Calculating accelerations} \label{sec:accelerations}
With the stellar positions determined over all epochs, we now require a transformation that relates the pixel positions to on-sky coordinates. As previously mentioned, we used a set of 91 reference stars to define the GC coordinate system to a level of ~0.7\,mas accuracy \citep{Gillessen2009}. Once pixel positions for all the  reference stars were obtained, we could determine a polynomial best fit that could convert pixel positions to right ascension (RA) and declination (Dec.) for each reference star at the relevant epoch. This was done using a least-squares fit and defining cubic transformations for both RA and Dec. individually. Applying the transformations to the stars detected by the source finder yields their RA and Dec. coordinates as a function of time. We then fit the positions for the star over all epochs it was observed using a second-order polynomial to determine the proper motions of the stars in RA and Dec.

In order to define the significance and direction of the accelerations detected for these stars, it is beneficial to rotate the coordinate system such that the motion of the stars is decomposed into its radial and tangential components. Similarly to the previous polynomial fit, we fit a second-order polynomial using \texttt{numpy.polyfit} \citep{numpypolyfit} to the motion as a function of time in this rotated coordinate system such that the results have the form
\begin{equation}
\begin{split}
    &\frac{1}{2}(a_{\mathrm{radial}} \pm \Delta a_{\mathrm{radial}})\,(t-t_{ref})^2 +\\
    &(v_{\mathrm{radial}} \pm \Delta v_{\mathrm{radial}})\,(t-t_{ref}) + (p_{\mathrm{radial}} \pm \Delta p_{\mathrm{radial}}),
\end{split}
\end{equation}
where $a_{\mathrm{radial}}$ is the acceleration in the radial component, $v_{\mathrm{radial}}$ is the velocity in radial component, $p_{\mathrm{radial}}$ is the radial component position of the star and the relative $\Delta$ terms refer to their respective errors, which are determined from the diagonal elements of the covariance matrix. The use of the $(t-t_{ref})$ term is due to the fact that the epochs are measured in years since 2010 (which is the approximate mid-point of the epoch list). The results are of the same form for the motion in the tangential direction. We used the polynomial fits to calculate the tangential and radial accelerations for each star and determined the significance of these accelerations. The significance of the radial acceleration, as well as its size and sign, indicates whether the motion of the star is dictated by the central SMBH. With this, we could quickly determine which stars show significant accelerations toward Sgr~A* as it is simply a case of examining the second-order term and its error. This acceleration term must be significant in the radial direction and consistent with zero in the tangential direction. The accelerations of stars measured to be significant in the tangential direction are an indicator of the confusion for these measurements, as there is no known body in the GC that would govern the stars to move with this type of significant tangential motion \citep{GRAVITYmassdist2021}.

\begin{figure}[h!]
    \centering
    \includegraphics[width=.45\textwidth]{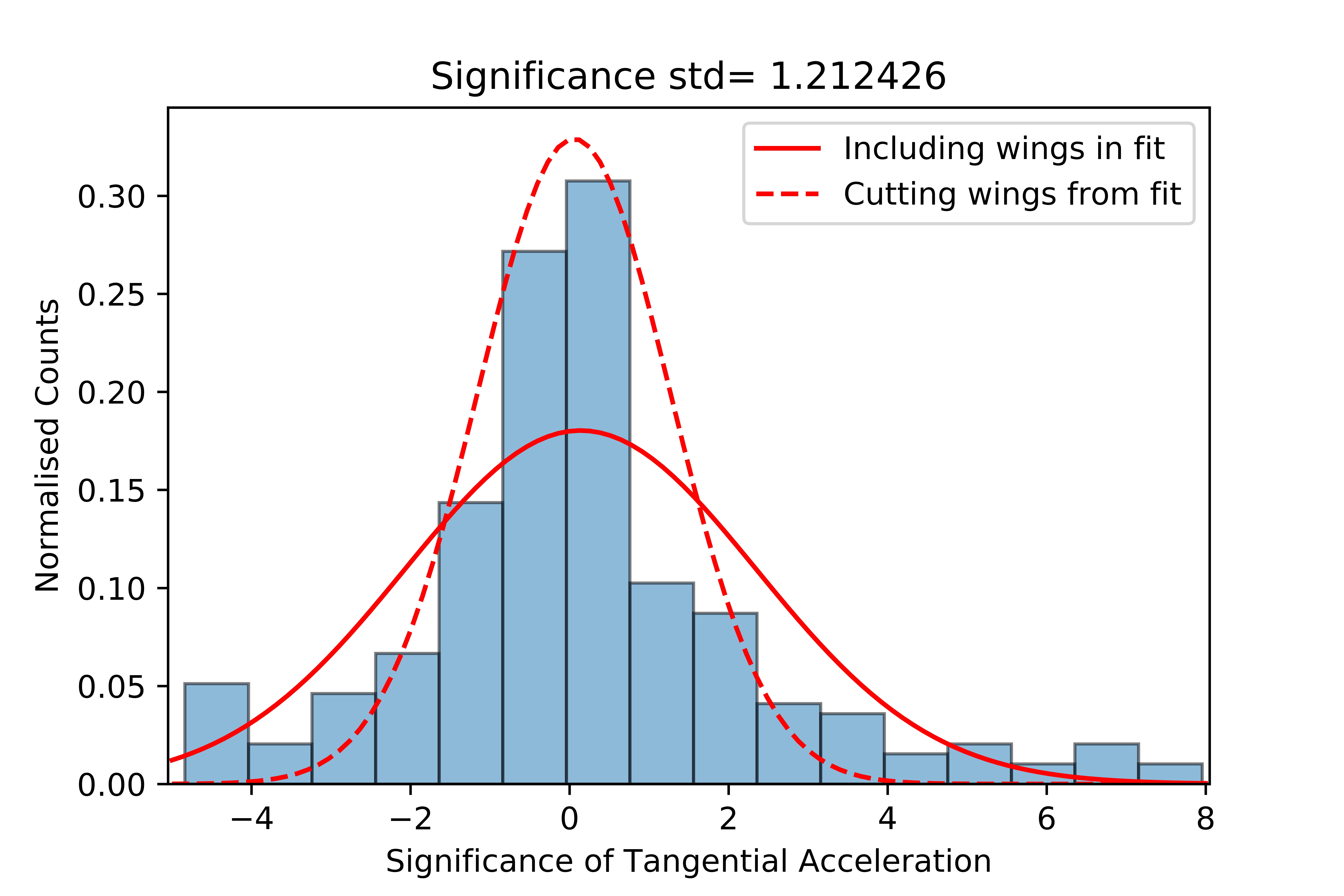}
    \caption{Histogram illustrating the distribution of tangential acceleration significance within the sample of stars examined. The histogram is normalized such that the area under the curve is equal to 1. The tangential significance for each star is calculated from Eq. \ref{eq:tangentialsignificance}. Two Gaussian fits to the histogram are plotted. The first includes the wings of the histogram when finding a best fit (\textit{solid curve}) that is dominated by sources with nonphysical accelerations, while the second excludes the data located in the wings of the histogram (\textit{dashed curve}). Therefore, the standard deviation for the curve, which excludes the wings, was calculated and used to normalize the radial acceleration significance. In this case, the standard deviation is found to be $\sim$\,1.212.}
    \label{fig:tangsig}
\end{figure}
Next, we defined a condition to determine when the radial acceleration is considered sufficiently significant. We did this by calculating the significance in both the radial and tangential directions for each source. First the tangential significance was calculated for each star from the following:
\begin{equation}
    \label{eq:tangentialsignificance}
   \sigma_{\mathrm{tangential}} = \frac{1}{\sqrt{\chi^2_{r}}} \frac{a_{\mathrm{tangential}}}{\Delta\,a_{\mathrm{tangential}}},
\end{equation}
where $\chi^2_{r}$ is the reduced chi-square of the second-order polynomial fit to the tangential component of the data, $a_{\mathrm{tangential}}$ is the coefficient from the second-order fit corresponding to the acceleration and $\Delta\,a_{\mathrm{tangential}}$ is the error. A histogram of the tangential acceleration significance for all sources was fit with two Gaussian profiles as shown in Fig.\index{Fig} \ref{fig:tangsig}. The solid red line shows the Gaussian that describes the data in full. However, we find that erroneous identifications and nonphysical accelerations dominate the wings of the histogram. For this reason, the second fit shown by the dashed red line excludes the wings and provides a more accurate Gaussian fit for the peak of the distribution that contains the relevant data. The standard deviation of this Gaussian distribution is measured to be $\sim 1.212$ and corresponds to the width of the dashed fit. This standard deviation was used to normalize the radial significance, namely,
\begin{equation}
   \sigma_{\mathrm{radial}} = \frac{1}{\sqrt{\chi^2_{r}} \sigma_{\mathrm{tangential}}} \frac{a_{\mathrm{radial}}}{{\Delta\,a_{\mathrm{radial}}}},
\end{equation}
where $\sigma_{\mathrm{tangential}}$ now refers to the standard deviation of the distribution shown in Fig. \ref{fig:tangsig}. The result is an adjusted significance that we used to define a list of stars that have a radial acceleration due to Sgr A*. 

Additionally, we excluded sources that exhibit an apparently nonphysical acceleration. We did this by calculating the maximally possible acceleration that is reached assuming $z=0$, where $z$ is the line-of-sight distance to the star:
\begin{equation}
\label{eq:Accelcondition}
   a_{(z=0)} =-\frac{G M_{\bullet}}{r_{2D}^2}.
\end{equation}
Here, $G$ is the gravitational constant, $M_{\bullet}$ is the mass of Sgr~A* (here taken to be $4.3 \times 10^{6}\, \mathrm{M_{\odot}}$ \citep{GRAVITYmassdist2021}) and $r_{2D}$ is the projected distance of the source from Sgr~A*. If the second-order fit produced a coefficient that corresponds to a greater acceleration than this maximum, the acceleration is nonphysical and this star was rejected from the list. In this case, the number of stars exhibiting nonphysical accelerations increases as we look in rings of increasing separations from Sgr A*. This demonstrates that the reasons for these nonphysical accelerations are largely due to the source finder misidentifying a star after a confusion event. Additionally, as the projected distance from Sgr~A* increases there are increased gaps between epochs where stars are detected due to the source locations not being recovered by the source finder. The stars further out are often fainter and depending on the pointing of the telescope, may be excluded from the FOV for some epochs. These are challenging conditions for the source finder that can lead to misidentifications that result in accelerations that exceed physical limits, sometimes by full orders of magnitude. In total, of the 677 stars that have negative accelerations, only 168 stars (approximately $25\%$) are not excluded due to nonphysical accelerations. Stars that exhibit physical accelerations are not likely to have their accelerations falsely increased by a similar effect, as inspection of the source positions in RA and Dec. would clearly reveal if source misidentifications have occurred.

Extending the use of the Lucy-Richardson algorithm to be applied to faint stars beyond separations of 5 arcseconds could improve the contrast between the sources and the background, thus improving the positions determined by the source finder, and in turn, reducing the number of stars featuring nonphysical accelerations.

We must now define our final list of stars with significant and physical accelerations due to Sgr A*. For this to be the case, both the acceleration and the significance calculated in the radial direction must be negative. For this reason, we set the condition such that $\sigma_{\mathrm{radial}} \leq -3$. For stars with a physical acceleration that does not meet this condition, continued monitoring may help improve the significance of their radial acceleration in the future.

With all of these conditions considered, in total, 54 stars within 2 to 7 arcseconds from Sgr A* are found to exhibit significant radial accelerations. These stars are key to understanding the dynamics resulting from the potential of Sgr A* and are, therefore, the focus of the remainder of this work. These stars are summarized in Table \ref{tab:idandradialsig}.

With this final list of stars that experience a significant radial acceleration, we can perform further analysis to determine more information about their motion within the GC. Hereafter, ``radial acceleration" will be referred to as ``astrometric acceleration" to avoid confusion with the traditional ``radial velocity" term referring to the $v_z$ component of the motion. The polynomial fits in both RA and Dec. for all 54 of these stars are provided in Tables \ref{tab:significantstarspolyfitsinner} and \ref{tab:significantstarspolyfitsouter}. 
\begin{table}[h!]
    \centering
    \caption{Significance of the stars exhibiting physical astrometric acceleration toward Sgr~A*}
    \begin{tabular}{cccccc}  \hline \hline
    ID & $\sigma_{\mathrm{radial}}$& ID & $\sigma_{\mathrm{radial}}$ & ID& $\sigma_{\mathrm{radial}}$\\ 
    &&(con't) & (con't) & (con't) & (con't) \\\hline
    3-3 & -4.101 & 3-89 & -7.86 & 4-207  & -6.188  \\
    3-4 & -10.336 & 3-91 & -7.272 & 4-241 & -4.03 \\
    3-7 & -6.329 & 3-103 & -6.322 & 3-190 & -10.614  \\
    3-9 & -7.101 & 3-104 & -22.82 & 3-244 & -13.141 \\
    3-17 & -7.866 & 3-105 & -25.614 & 4-19 & -5.359\\
    3-18 & -6.995 & 3-111 & -14.768 & 4-29 & -4.448\\
    3-19 & -24.749 & 3-114 & -3.004 &4-36 & -10.546\\
    3-20 & -16.152 &3-134 & -11.081 & 3-81 & -15.221  \\
    3-27 & -7.455 & 3-142 & -5.486 & 3-143 & -4.131 \\
    3-40 & -9.061 & 3-158 & -7.371 & 4-46 & -7.398\\
    3-47 & -9.286 &3-161 & -3.149   &  4-56 & -5.833   \\
    3-52 & -25.555 & 4-311 & -3.147 &  4-62 & -4.814  \\
    3-53 & -5.577 & 4-326 & -3.097 &  4-64 & -10.073 \\
    3-58 & -22.379 & 4-329 & -5.814 &4-79 & -3.71 \\
    3-72 & -8.801 & 4-272 & -3.01 &4-83 & -4.804  \\
    3-74 & -10.848 &4-211 & -3.307  &  4-109 & -3.28\\
    3-84 & -5.631  &5-55 & -4.071 &  4-120 & -6.54\\
    3-87 & -26.002 & 5-139 & -5.568 & 5-246 & -7.666 \\\hline
   \end{tabular}
    \label{tab:idandradialsig}
    \tablefoot{IDs of the 54 stars exhibiting physical astrometric acceleration toward Sgr~A* with a normalized significance $< -3$, corresponding to the stars selected for further investigation.}
\end{table}
\section{Determining stellar orbits} \label{sec:orbitalelements}
All 54 of the stars featuring significant astrometric accelerations in Table \ref{tab:idandradialsig} were then classified as ``5D constrained'' stars. This means that, of the six phase-space coordinates required to determine the six orbital elements, five were known. The known elements included the on-sky position of the star ($x$, $y$) and the proper motion of the star ($v_x$, $v_y$). Additionally, by determining the acceleration from fitting the on-sky orbit trace in Sect. \ref{sec:accelerations}, we could determine the absolute value of the $z$ coordinate for each star, given by
\begin{equation} \label{eq:absz}
|z|= \sqrt{\left|\left(\frac{-G M_{\bullet}r_{2D}}{a_{2D}}\right)^{2/3}-r^2_{2D}\right|}, 
\end{equation}
where $a_{2D}$ is the 2D acceleration term, $a_{2D}=~\sqrt{a_x^2~+~a_y^2}$, and similarly, $r_{2D}$ is the 2D position vector.

The missing phase-space coordinate was then the velocity in the $z$ coordinate, or the radial velocity of the star. We used two methods to extract radial velocity information for the 54 stars demonstrating significant acceleration toward Sgr~A* and determine their orbital solutions. Sect.~\ref{subsec:measradvel} details the first method involving the determination of radial velocities from spectroscopic observations. For stars where we could not obtain a reliable spectrum, we applied Monte Carlo sampling in radial velocity to determine a range of physically viable orbital solutions. This method is discussed in Sect.~\ref{subsec:dicing}.
\subsection{Measured radial velocities from SINFONI data} \label{subsec:measradvel}
To obtain spectra for the stars listed in Table \ref{tab:idandradialsig}, we used datacubes from the ESO's decommissioned SINFONI instrument. Stars were identified in the IR image via cross-comparison to NACO images with the nearest epoch to the SINFONI datacube. Once identified, we extracted their spectra. From the stellar spectra, we could then classify the stars as either spectroscopically old or young depending on the characteristic features present in the spectrum of the star. This process is explained in detail by \cite{Sebyoungstars} but can be summarized as follows: the Bracket-$\gamma$ (Br-$\gamma$) line located at $\sim 2.166 \,\mu$m or the, often less discernible, helium absorption line at $\sim 2.113\, \mu$m is considered as a criterion for young stars and the CO band heads that start at $\sim 2.3\, \mu$m for old stars.

In some cases the noise level in the spectrum was too high and we could not confidently extract a line from the spectrum. We also utilized line maps for stars that had faint absorption lines to confirm their origin was consistent with the location of the source and the expected full width at half maximum source diameter. 

The radial velocities were then determined from the measurement of the midpoint of the relevant line and the Doppler equation. We then corrected these to the local standard of rest. The results for the stars that had detected spectra are contained in Tables \ref{tab:orbitalelementsposz} and \ref{tab:orbitalelementsnegz}. The errors for the radial velocities measured with SINFONI are large due to low signal-to-noise ratios and intrinsically large line widths (especially for faint sources). This means the resulting orbital elements have errors dominated by this measurement. 

In total, we could determine radial velocities for 20 of the 54 stars, either by comparison to previously studied catalogs (e.g., \cite{Yelda2014}, \cite{Sebyoungstars}) or by extracting the radial velocities from SINFONI spectra. Of the 20 stars, three stars already have orbital elements published in previous works, namely \mbox{3-74}, \mbox{3-103,} and \mbox{3-134,} which are otherwise known as the  reference stars R74, R30, and R34, respectively. These stars were left in the sample to serve as a control group. A total of ten of the stars for which radial velocities could be determined were found to be classified as old stars, while the remaining ten were determined to be young stars. Of the young stars, three had not been identified previously. For all stars that have measured radial velocities, we calculated the six orbital elements using all six phase-space coordinates up to the degeneracy induced by the unknown sign of the $z$ coordinate. 

\begin{figure*}[h!]
  \centering
    {
      \includegraphics[width=.99\textwidth]{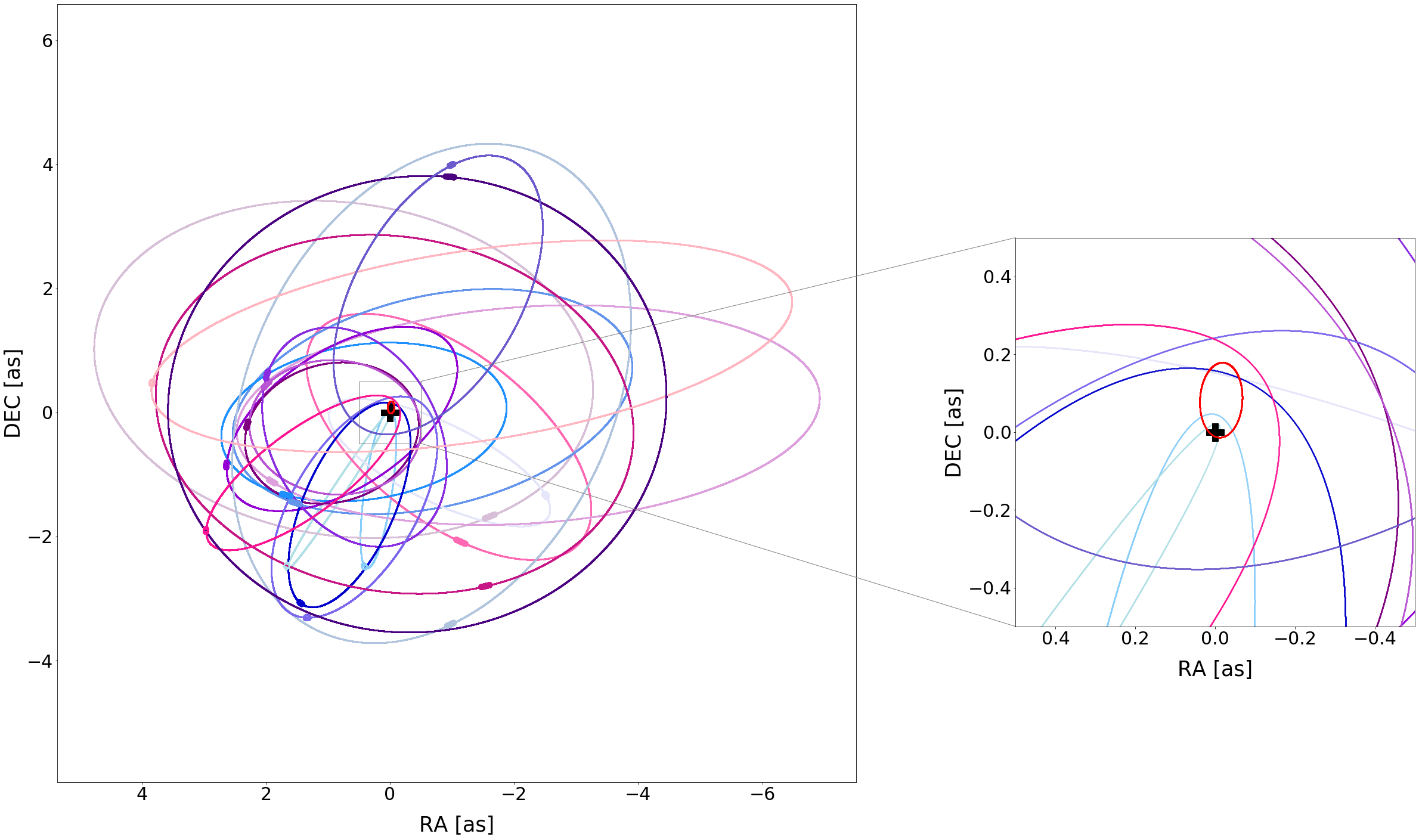}
      \includegraphics[width=.99\textwidth]{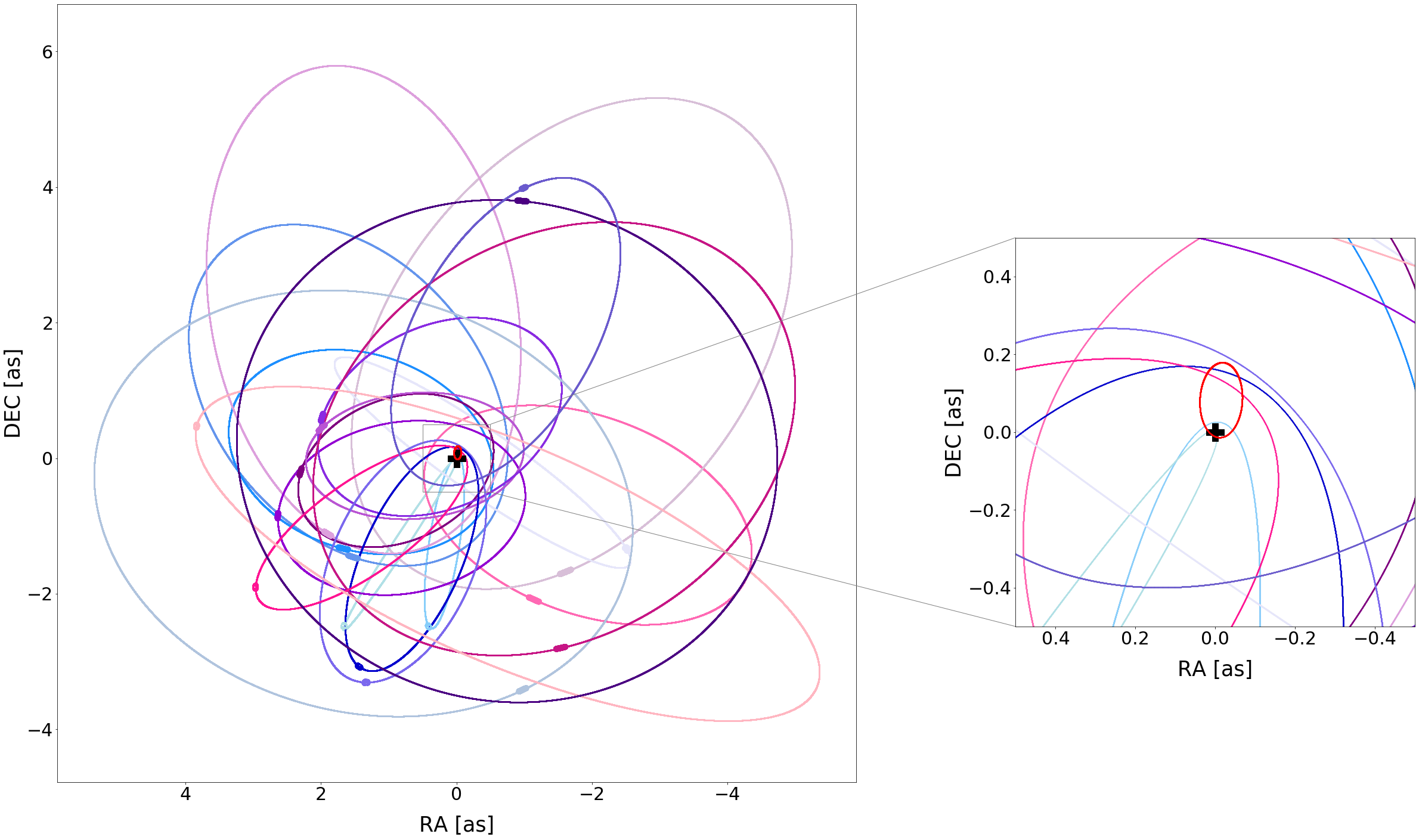}
    }
 \caption{All stars with full orbital solutions determined using negative $z$ coordinates (\textit{upper}) and positive $z$ coordinates (\textit{lower}). The black plus sign denotes the location of Sgr~A* at the (0, 0) coordinate, and NACO data points are marked by the filled circular points along each orbit. We also include the orbit of the S star S2 in red to provide a sense of scale.}
 \label{fig:allorbitscombo}
\end{figure*}

\begin{table}[h!]
    \centering
    \caption{Time required to observe a change in radial velocity of $50\,$km/s for stars with measurable radial velocities.}
    \begin{tabular}{cccc}\hline\hline
         ID  &  $t_{\mathrm{elapse}}$ [yr] & ID (con't)  &  $t_{\mathrm{elapse}}$ [yr] (con't)\\\hline
         3-19 & 0.005 & 3-158 & 9.072\\
         4-120 & 0.101 & 3-134 & 9.262\\
         3-20 & 0.137& 3-89 & 16.039\\
         4-56 & 0.227 & 4-79 & 25.851\\
         4-36 & 0.255 & 3-74 & 26.776\\
         3-53 & 0.847 & 3-111 & 38.016\\
         3-91 & 2.243 & 4-326 & 68.217 \\
         5-246 & 3.849 & 3-84 & 103.323\\
         3-103 & 5.433 & 4-19 & 145.74\\
         3-161 & 5.641 & 4-207 & 357.304\\
         \hline
    \end{tabular}
    \label{tab:elapsetimeto50kmchange}
    \tablefoot{Time required to observe a change in radial velocity of $50\,$km/s for the 20 stars that show significant astrometric accelerations and have measurable radial velocities. Measuring the change in radial velocity of a star could be used to determine the correct sign for the $z$ coordinate and lift the degeneracy of their orbital elements.}
\end{table}
It is often the convention to express orbital elements as the Keplerian parameters, which are: the semimajor axis, $a$, the eccentricity, $e$, the inclination, $i$, the argument of periapsis, $\omega$, the longitude of the ascending node, $\Omega$, and the epoch of the pericenter passage, $t_p$. The resulting orbital solutions for the 20 stars with radial velocity measurements are contained in Tables \ref{tab:orbitalelementsposz} and \ref{tab:orbitalelementsnegz}. We also visualized these orbits by plotting them together as shown in Fig. \ref{fig:allorbitscombo}. With the addition of these stars at larger separations from Sgr~A*, we have increased the number of known orbits within the GC by $\sim40\,\%$.

If it were possible to observe a change in the radial velocity for these stars, we would be able to determine the correct sign for the z coordinate. Additionally, this would mean seven phase space coordinates are known while only six degrees of freedom apply to the calculation of the orbital elements for each star. With the additional phase-space coordinate, these stars could be used to constrain the potential of Sgr~A* and, therefore, the SMBH mass. In this way, a prediction of when it would be possible to observe a significant change in radial velocity for each star is valuable. If it is assumed that a change of $\Delta\,v_{rad} = 50\,$km/s would be robustly detectable, then it follows that the time that must elapse from now until such a change is observable is given by dividing the change in radial velocity by the radial acceleration, $a_z$, of the star. The choice of $50\,$km/s is due to the fact that this change in radial velocity will be detectable using the Enhanced Resolution Imager and Spectrograph (ERIS), which has begun commissioning at the ESO VLT in 2022. The values obtained are provided for each star in Table \ref{tab:elapsetimeto50kmchange}.
\subsection{Monte Carlo sampling over flat radial velocity prior} \label{subsec:dicing}
For stars that do not have measurable radial velocities, we used Monte Carlo sampling over a flat prior for the radial velocity to determine possible orbital solutions. The method follows closely that outlined by \cite{warpedBartko2009} and \cite{Lu_2009}. However, for the stars studied in \cite{warpedBartko2009} and \cite{Lu_2009}, the missing phase-space coordinate was the $z$ coordinate, which has been determined in this work by detecting the curvature in the on-sky orbit trace. 

In this case, the missing phase-space coordinate, here $v_z$, was subject to a series of assumptions. Firstly, we assumed that the star is gravitationally bound to Sgr~A*. As such, a maximum velocity is readily calculated as the escape velocity of the star. If the maximum velocity vector is then $v_{esc} = v_{max} = \sqrt{v_x^2+v_y^2+v_{z,\mathrm{max}}^2}$, the absolute value of $v_{z, \mathrm{max}}$ is then determined from\begin{equation}
    |v_{z,\mathrm{max}}| = \sqrt{\left|\frac{2\,G\,M_{\bullet}}{r}-v_x^2-v_y^2\right|}.
\end{equation}
In this case, we considered all solutions in the range $v_z~\in~[-v_{esc}, v_{esc}]$ by generating 20000 velocities between these two values from a flat distribution. This was performed twice to accommodate the fact that it is again not possible to determine whether the $z$ coordinate is positive or negative, which results in each star having 40000 physically possible orbital solutions. We plotted these solutions onto a spherical Hammer-Aitoff projection map with axes corresponding to the longitude of the ascending node, $\Omega$, and the inclination, $i$. The result is two curved traces representing possible orbital solutions for each star on the map. 

\begin{figure}[h]
  \centering
    {
      \includegraphics[width=.45\textwidth]{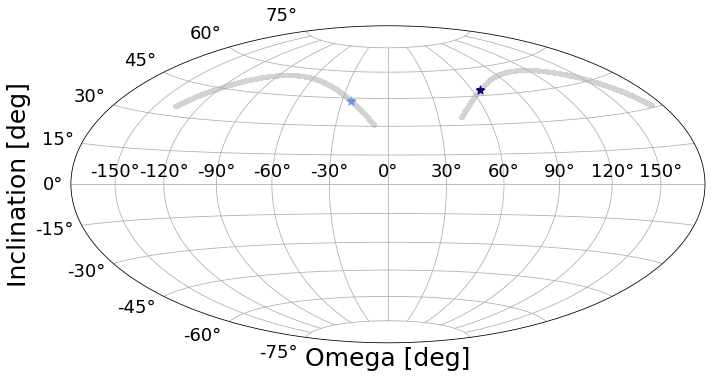}
    }
     \caption{Projected $\Omega$ vs. $i$ traces determined via Monte Carlo sampling in radial velocity for the star 4-19. The ranges of the determined physically possible solutions are shown by the gray points, and the exact orbital solutions determined via radial velocity measurements from SINFONI data are indicated by the blue star markers for both the positive $z$ (\textit{navy}) and negative $z$ (\textit{light blue}) solutions. The errors in [$x,~y,~z,~v_x,~v_y$] are taken into account by sampling these values from a Gaussian distribution, but the resulting error bars are excluded from the figure. This consideration results in an uncertainty in $\Omega$ for the positive $z$ solution of d$\Omega = 20.7^{\circ}$ and d$\Omega =15.8^{\circ}$ for the negative $z$ solution; for $i$ it is given as d$i = 6.7^{\circ}$ and d$i = 7.4^{\circ}$ for positive and negative $z$ solutions, respectively. }
  \label{fig:examplemap}
\end{figure}

An example of such a map of solutions is provided in Fig. \ref{fig:examplemap}. The coordinates used to produce Fig. \ref{fig:examplemap} are those used in \cite{Sebyoungstars}. However, previous works (see, e.g., \citealt{2disksPaumard2006,Gillessen2009,Gillessen2017}) use a slightly different definition, which is compared to the coordinates used in this work in Appendix \ref{AP: othercoordhamaps}. 

We could also determine the direction of rotation of the stellar orbit from these projections. More specifically, if a star has solutions located on the northern hemisphere of the map, then the star orbits clockwise, and if located in the southern hemisphere, the orbit is counterclockwise.
\section{The Galactic Center environment} \label{sec:gcenvironment}
We can use the results discussed in the previous section to further understand the environment of the nuclear star cluster in the GC at these separations from Sgr~A*. More specifically, in Sect. \ref{subsec:diskmembers} we use the Hammer-Aitoff projection plots to compare previous works that looked for overdensities of stars in these $(\Omega,i)$ coordinates to indicate which stars could belong to disks or features in the GC environment (e.g., \citealt{warpedBartko2009,2disksPaumard2006,Yelda2014,Sebyoungstars}). Once disk membership is assumed for a star, we use this information to examine the mass enclosed within the orbits of these stars by removing the dependence of the black hole mass from the stellar $z$ coordinate. We examine this in Sect. \ref{subsec:massdist}. Finally, by comparison to a simulated isotropic cluster, we investigate further structural information via trends in the orbits of the stars studied in Sect. \ref{subsec:isoclus}.
\subsection{Comparison to known disk structures} \label{subsec:diskmembers}
\begin{figure}[h!]
  \centering
    {
      \includegraphics[width=.5\textwidth]{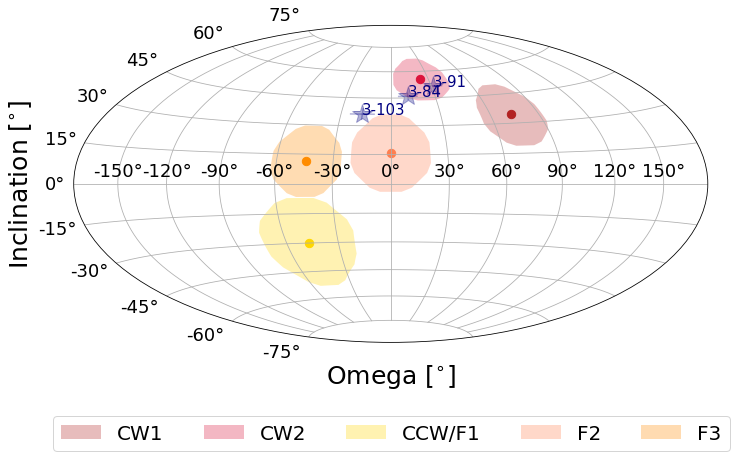}
    }
     \caption{Projected $\Omega$ vs. $i$ orbital solutions for the three young stars that have known orbital elements that correspond to a known GC disk feature (within error bars). All features are marked here by ellipses (see, e.g., \citealt{warpedBartko2009,Yelda2014,Sebyoungstars}). Disk features are as calculated for Fig.~7 of \cite{Sebyoungstars} (hereafter, vF22).}
  \label{fig:knownstarspoints}
\end{figure}

Fig. \ref{fig:knownstarspoints} shows the $\Omega-i$ projection of spectroscopically young stars with known orbital elements corresponding to the solutions found in Sect. \ref{subsec:measradvel} that coincide with known overdensities in the GC environment. These observable disk features in the GC are those calculated in \cite{Sebyoungstars} (hereafter, vF22) and are visually summarized as ellipses in Fig. \ref{fig:knownstarspoints}. The sample studied in vF22 contained stars with separations from Sgr~A* up to 32 arcseconds and, therefore, contains the regions studied in this work. As such, this map allows for a comparison between the stellar orbital solutions calculated in Sect. \ref{subsec:measradvel} and the known disk features in the GC. In this case, only stars that were identified as young stars using stellar spectra are considered since these features are typically populated only by young stars (see, e.g., \cite{2disksPaumard2006,Bartko2010,Genzel_2010_gcreview}; vF22). The features include the inner clockwise disk (CW1), the outer clockwise disk (CW2), the counterclockwise disk/filament (CCW/F1), as well as outer filament 2 (F2) and outer filament 3 (F3). In this case, only stellar orbital solutions that are consistent (within errors) with the values of $i$ and $\Omega$ representing disk features are plotted. As such, Fig.~\ref{fig:knownstarspoints} indicates that three of the stars in this work could belong to either CW2 or F2. We can use this information to predict the sign of the $z$ coordinate by assuming the correct sign of $z$ corresponds to the solution that results in the star belonging to a disk feature. This, of course, self-amplifies the disk features.\ However, it can be argued that this is a reasonable assumption since these features indicate overdensities in the stellar population in these coordinates. If a star has both positive and negative $z$ solutions that are consistent with a feature, we cannot use this information to lift the orbital degeneracy. For this reason, only stars with a single $z$-solution consistent with a disk feature are listed in the summary of possible disk members provided in Table \ref{tab:diskstars}. The results are generally consistent with prior knowledge of the structure of the nuclear star cluster. It was determined by vF22 that particular features dominate at different separations from Sgr~A*. From the stars studied here, it is clear that mostly stars within the first 2'' to 3'' are found on the outer clockwise disk. Unfortunately, Fig. 4 of vF22 also indicates that the slice with features at the lowest significance is from 4'' to 8'', meaning a smaller fraction of stars at these separations lie on features. Further, no single disk feature seems to dominate the density map at these separations. Since this largely represents the stars with larger separations studied in this work, it is not surprising that eight of the young stars with determined orbital parameters do not appear to be consistent with any features. Additionally, it could be argued that the source 3-103 is unlikely to belong to the Outer Filament 2, as this feature is most prominent at larger separations. Generally, the results of this study are consistent with the understanding of the nuclear star cluster presented in vF22.

\begin{figure}[h!]
  \centering
    {
      \includegraphics[width=.48\textwidth]{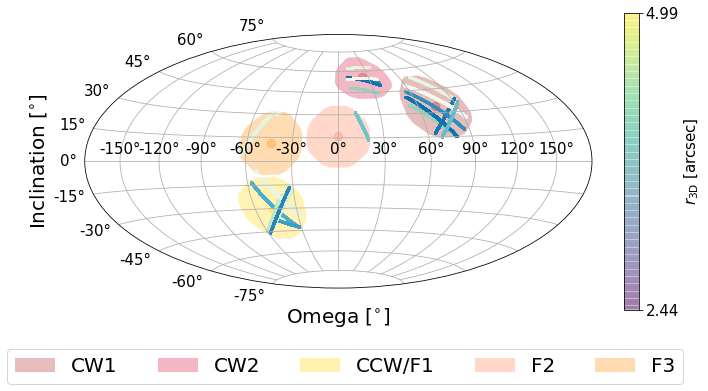}
    }
     \caption{Projected $\Omega$ vs. $i$ traces determined via Monte Carlo sampling over a flat distribution in radial velocity for all stars that have a subset of solutions that corresponds to at least one of the known GC disk features using only a single sign for the $z$ coordinate. The features are shown here by ellipses (see, e.g., \cite{warpedBartko2009,Yelda2014}; vF22). Disk features are as calculated for Fig. 7 of vF22. The color of the trace indicates the 3D position vector of the source given by $r_{\mathrm{3D}}= \sqrt{x^2+y^2+z^2}$. This illustrates that the stars with the lowest $r_{\mathrm{3D}}$ appear on the CW disk. This is consistent with the findings of vF22.}
\label{fig:unknowndicingmaps}
\end{figure}

 \begin{table}
    \centering
     \caption{Young stars with an orbital solution consistent with a disk feature.}
    \begin{tabular}{ccccccc}  \hline  \hline
    ID & +z/-z & $\Omega [^{\circ}]$ & $d\Omega [^{\circ}]$ & $i [^{\circ}]$ & $di [^{\circ}]$ & Disk Feature \\ \hline
    3-84 & + & 22 & 24.8 & 47.9 & 4.4 & CW2 \\
    3-91 & + & 17.9 & 29.5 & 47.6 & 5.0 & CW2 \\
    3-103 & + & 3.3 & 26.0 & 43.9 & 5.4 & F2\\
 \hline
 \end{tabular}
 \tablefoot{Spectroscopically young stars with full orbital solutions that have either positive or negative $z$ orbital solutions consistent with disk features presented in \cite{2disksPaumard2006,Lu_2009,Yelda2014} and vF22. Here, stars that appear to be consistent with a disk feature for both positive and negative $z$ solutions are excluded.}
 \label{tab:diskstars}
\end{table}
We can use a similar analysis for stars that did not have measurable radial velocities to reveal which ranges of solutions appear to be consistent with the GC features. Fig. \ref{fig:unknowndicingmaps} can be used to constrain the possible $z$ values for stars that have solutions corresponding to disk features. Here, if a star without a measurable spectral radial velocity had orbital solutions determined from the Monte Carlo sampling of radial velocities that coincide with any of the known GC features, they were plotted onto Fig. \ref{fig:unknowndicingmaps}. Again, we could use this information to predict the correct sign of the $z$ coordinate for these stars. Table \ref{tab:5Dfeaturesandz} provides a list of stars that only appear on disk features using one of the signs for the $z$ coordinate. A total of eight stars have solutions that correspond to disk features if a negative $z$ coordinate is used, while six with positive $z$ coordinates result in solutions lying on a disk features. For stars that list multiple features, there may be one feature that is more suitable. For example, for stars that have solutions consistent with both the clockwise disk and the outer filaments, the solutions corresponding to the clockwise disk are more reasonable. This is because the CW disk is the dominant feature for stars at the separations from Sgr~A* consistent with this sample, and the filaments are generally dominant for larger separations (vF22). For 20 stars, it was either the case that no solutions were consistent with disk features or that solutions corresponding to disk features were possible with both positive and negative $z$ coordinates.
\begin{table}
    \centering
    \caption{5D constrained stars that have orbital solutions consistent with disk features.}
    \begin{tabular}{cc|cc}  \hline  \hline
    \multicolumn{2}{c}{Negative $z$ Solutions} &\multicolumn{2}{c}{Positive $z$ Solutions}\\ \hline \hline
    ID & Disk Feature& ID & Disk Feature \\ \hline
     3-105 & CCW/F1 & 3-9 & CCW/F1\\ 
     3-142 & CW1,& 3-18 & CCW/F1\\
      & CW2 & 3-114 & CW2,\\
     3-143 & CCW/F1 &  & F3\\
     4-62 & F2 & 3-190 & CCW/F1,\\
     4-241 & CW1, &  & F3\\
      &  CW2, & 4-109 & CW1\\
      & F3 & 5-55 & CCW/F1\\ 
    4-272 & CW1 & & \\
     4-211&CW1,&& \\ 
      & CW2& &\\
     5-139 & CW1, & &\\
     & CW2\\
 \hline
\end{tabular}
\tablefoot{5D constrained stars that only appear on disk features if one sign for the $z$ coordinate is assumed. Listed are the features consistent with possible solutions for the star as well as the sign for the $z$ coordinate that produces these solutions. Results are determined from Fig. \ref{fig:unknowndicingmaps}. }
\label{tab:5Dfeaturesandz}
\end{table}
\subsection{Using disk membership to estimate the mass distribution} \label{subsec:massdist}
\begin{figure}[h!]
  \centering
        \includegraphics[width=0.44\textwidth]{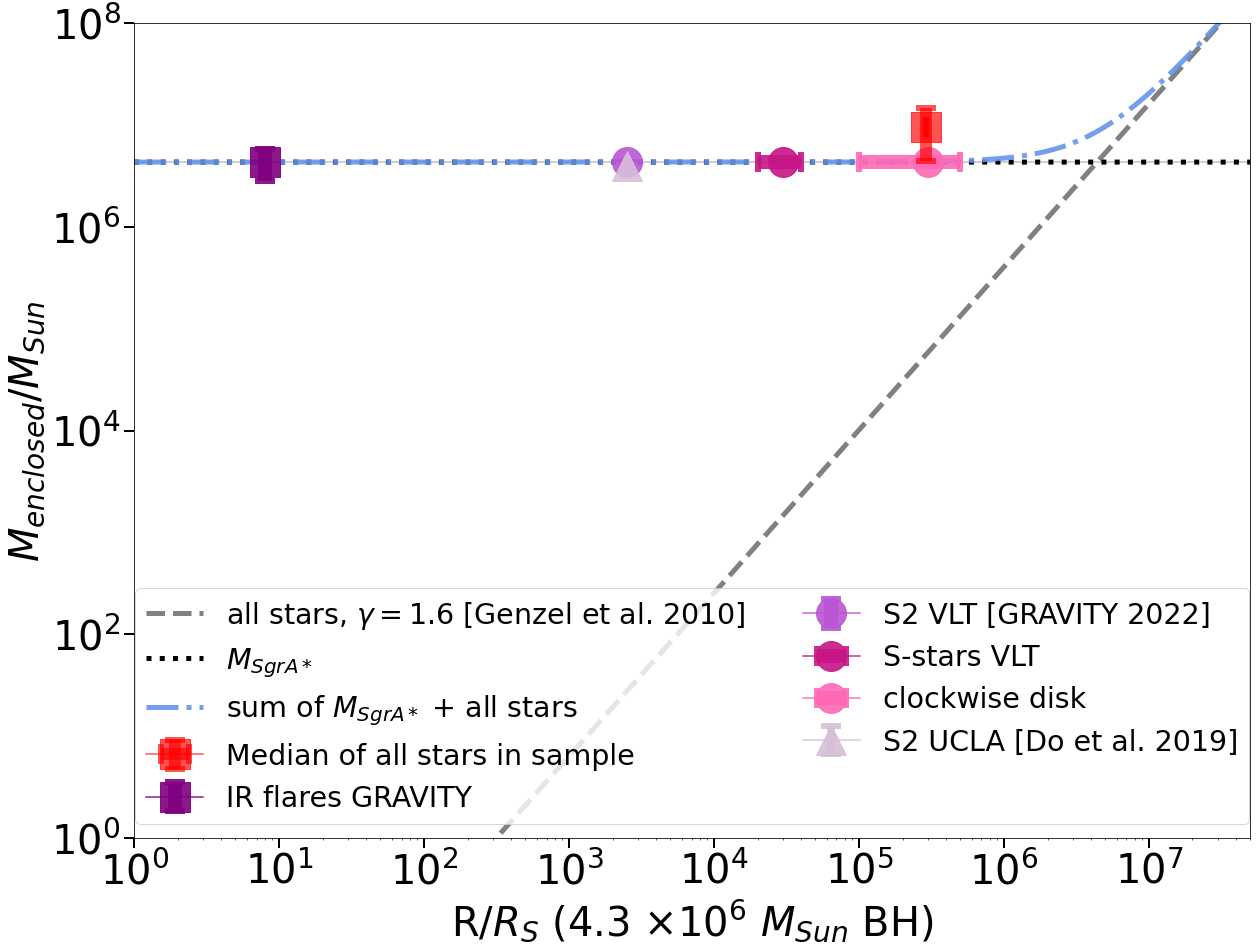}
     \caption{Distribution of mass contained within the orbits of various sources. The median of the stars that have orbital solutions consistent with disk features in this work is shown by the red square marker. In this case, since the $z$ values used are derived from the disk priors and we did not marginalize over priors, no formal error estimates for individual points can be provided. However, the uncertainty for the median of all the points in this work is estimated as the spread of the individual points. Also included are the values from the flare data \citep{2018gravityflares}, S2 from both \cite{Do2019} and \cite{GRAVITYmassdist2021}, multiple S stars as measured with the VLT \citep{GRAVITYmassdist2021}, and stars that are members of the clockwise disk \citep{GRAVITYmassdist2021}. In this work, the enclosed mass is consistent within errors with the clockwise disk value, meaning no additional mass is required to account for the mass contained in the orbits of these stars. The blue line shows the sum of $M_{\mathrm{Sgr~A*}}$ and the extended stellar mass distribution from previous works (e.g., \citealt{Ghez_2008, Gillessen2009, Genzel_2010_gcreview, talexander2017,schodel2018,baumgardt2018,GRAVITYmassdist2021}). }
     \label{fig:massdist}
   \end{figure}

The stars listed in Table \ref{tab:diskstars} and Table \ref{tab:5Dfeaturesandz} along with the assumption of disk membership can be used to further understand the distribution of mass located within their orbits. We did this using the longitude of the ascending node and inclination of the disk features consistent with their original orbital solutions. Using these $\Omega$ and $i$ values provides a novel method for calculating the $z$ coordinate for each star that does not directly rely on the use of the accelerations calculated from the on-sky orbit trace. This means the final $z$ coordinate will not depend on the mass of Sgr~A*, as is the case for the absolute value of $z$ calculated from Eq. \ref{eq:absz}. This was determined using
\begin{equation}\label{eq:znomsgr}
    z = \frac{x\,(\sin{i}\,\sin{\Omega}) + y\,(\sin{i}\,\cos{\Omega})}{-\cos{i}},
\end{equation}
where $x$ and $y$ are the stellar positions in Dec. and RA, respectively, and the selected $\Omega$ and $i$ coordinates are those that were determined in vF22 for the corresponding disk feature listed in Table \ref{tab:diskstars} and Table \ref{tab:5Dfeaturesandz}. 
We then calculated the enclosed mass within the orbit of the star by solving Eq. \ref{eq:absz} for $M_{\bullet}$. Fig. \ref{fig:massdist} shows the median of the resulting masses enclosed within the orbits of each star as a function of 3D distance. Again, for stars with known spectroscopic ages, only young stars were included in the calculation. In this case, since we are only demonstrating this method as a possible way to estimate the mass distribution at these separations, a simplified approach is employed. This means, no marginalization over the priors was used to determine uncertainties for the mass enclosed in individual stellar orbits. 

The enclosed mass calculated for stars in this work is consistent within its error with the enclosed mass of the clockwise disk stars in \cite{GRAVITYmassdist2021}. Since seven of the fourteen stars used in this calculation have solutions corresponding to either CW1 or CW2 and most of the stars are at separations consistent with the CW disk it is not surprising to see that the enclosed mass calculated here is consistent with this value. However, the points determined in this work do not further constrain the mass distribution since the required assumption of disk membership is heavily influenced by the uncertainties in radial velocity for stars with extracted spectral lines. For stars with unknown radial velocities, it is clear that the spread in possible orbital solutions also contributes to the error in the estimated mass enclosed within the potential orbits of these stars. This further highlights the importance of reducing the uncertainties in the stellar spectra used to calculate the radial velocities for these stars as well as needing better spectra for stars where radial velocities could not be determined. Future instrumentation with increased resolution promise to provide better estimates of disk membership and, thus, contribute to constraining the distribution of mass within the GC using stars with greater separations from Sgr~A*. 
\subsection{Comparison to a simulated isotropic cluster} \label{subsec:isoclus}
It is also useful to compare the studied sample of 54 stars with significant accelerations toward Sgr~A* to a simulated isotropic cluster to determine the nature of the nuclear star cluster at these separations. The procedure to calculate the six phase-space coordinates for each cluster member is detailed in \cite{genzel2003} and vF22, where first an isotropic distribution of orbits was determined, which we then converted to stellar positions and proper motions. The distributions representing the orbital elements for stars in the resulting isotropic cluster are shown in Fig. \ref{fig:isotropicdisthisto}.

In order to compare this simulated cluster to the data, we calculated the angular momentum statistic \citep[or $j$ statistic;][]{genzel2003} and the high-eccentricity statistic \citep[or $h$ statistic;][]{bstarsMadigan2014} for stars in the sample as well as the simulated isotropic cluster members. Both statistics were developed as a means to determine estimates for the orbital elements of stars located far enough away from Sgr A* that full orbital solutions could not be determined with less than a decade of observation. In this work, the $j$ and $h$ statistics are used only to serve as a powerful visual comparison between the simulated isotropic cluster and the sample of stars found to have significant astrometric accelerations toward Sgr~A*.

\begin{figure}[h!]
  \centering
    \hfill
    {
      \includegraphics[width=0.44\textwidth]{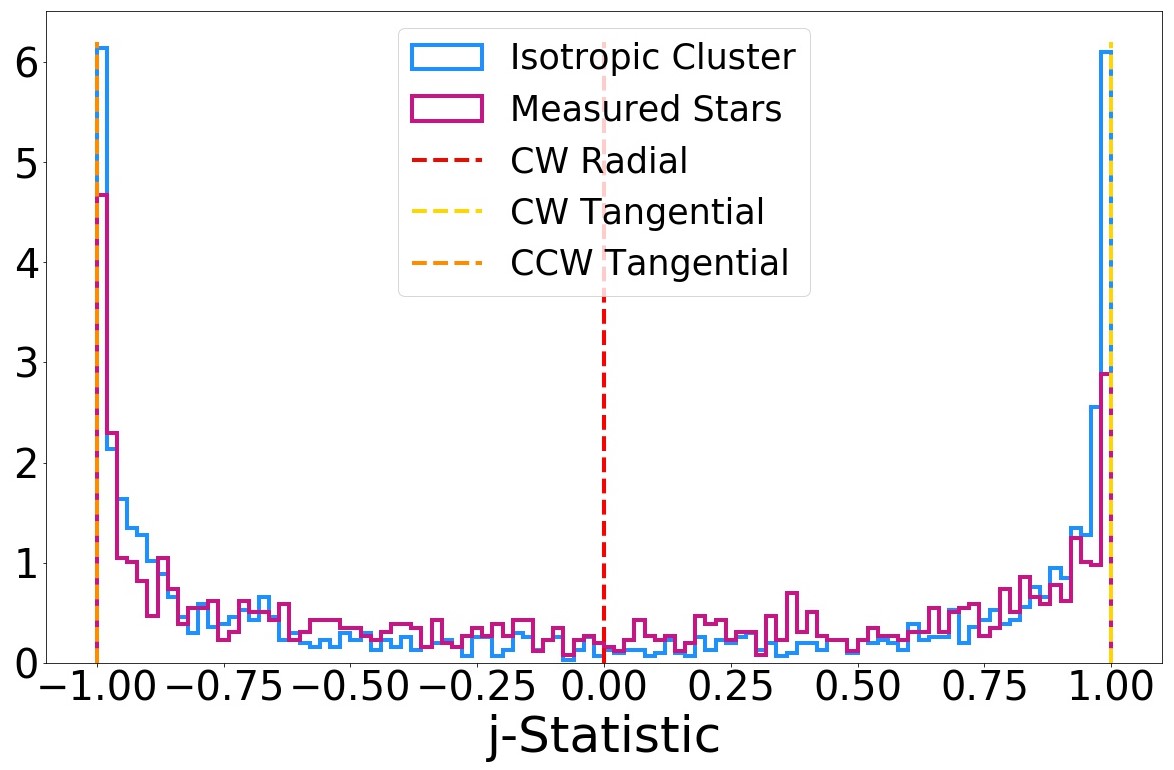}
 }
   \caption{Number of stars binned as a function of the calculated $j$ statistic for both the isotropic cluster (\textit{blue}) and observed stars (\textit{magenta}). Vertical lines indicate CW tangential (\textit{yellow}), CCW tangential (\textit{orange}), and CW radial (\textit{red}) on-sky projected orbits at 1, -1, and 0, respectively. These results suggest an overdensity of stars with CCW tangential orbits compared to the isotropic cluster that displays equal amounts of CW and CCW tangential orbits. To best model observations, all members of the isotropic cluster with separations of $< 2\,$arcseconds from Sgr~A* are discarded before plotting. }
    \label{fig:bothj}
\end{figure}

Fig. \ref{fig:bothj} shows the comparison between the observed stars with significant accelerations and the isotropic cluster via histograms of calculated values of the $j$ statistic. Here, the values of $j$ = -1, 0 and 1 are shown by the vertical lines and correspond to CCW tangential, CW radial and CW tangential orbits, respectively. Interestingly, it appears that the observed stars exhibit a larger number of CCW tangential on-sky orbits than CW tangential orbits, whereas the isotropic cluster exhibits an equal fraction of both. We also see that a larger portion of the measured stars exhibit $j$ values consistent with CW radial orbits as there is a surplus of stars that have $j\sim0$ when compared to the isotropic cluster. More quantitatively, when a Kolmogorov-Smirnov goodness-of-fit test (KS test) is performed using \texttt{scipy.stats.ks\_2samp} \citep{scipy.stats.ks} to compare the calculated $j$ values for both the observed stars and the simulated isotropic cluster, a KS score of $\sim 0.12$ is returned. This corresponds to a p value of $\sim 9.4 \times 10^{-9}$, meaning we can reject the null hypothesis in favor of the scenario where these values are drawn from different distributions.

\begin{figure}[h!]
  \centering
    \hfill
    {
      \includegraphics[width=0.43\textwidth]{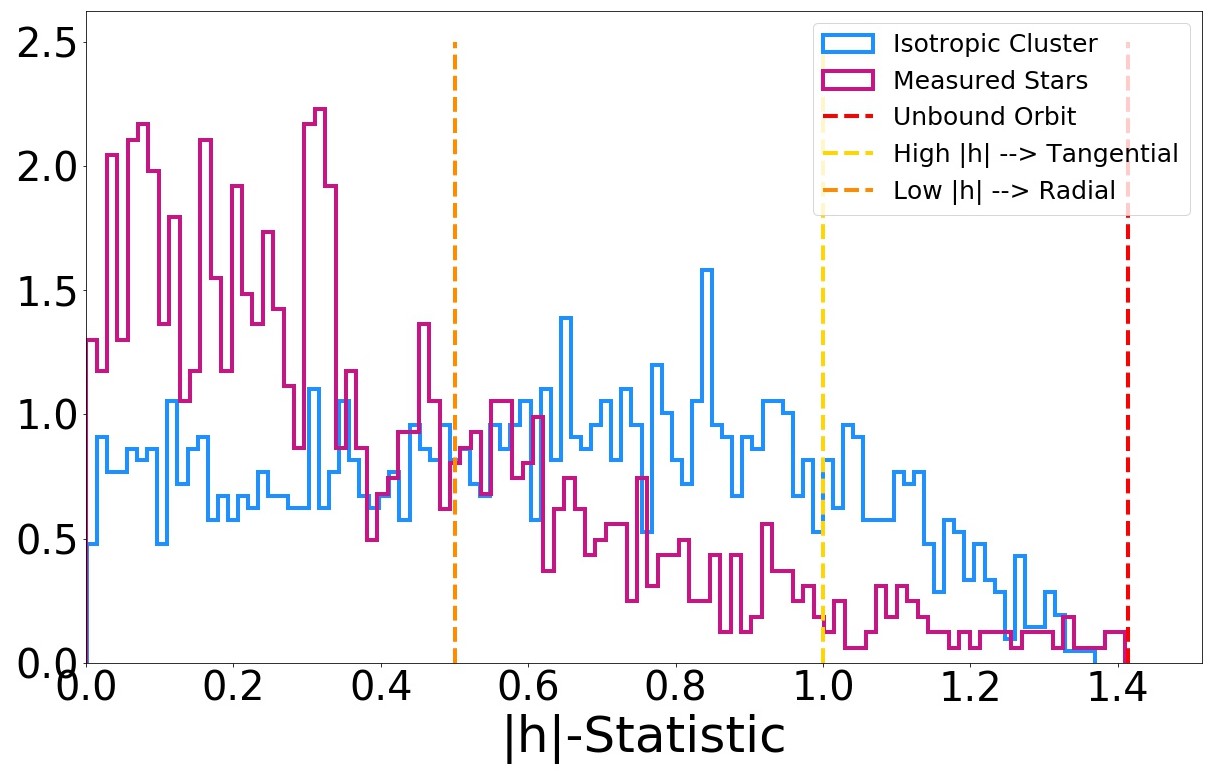}
 }
   \caption{Number of stars binned as a function of the calculated $|h|$ statistic for both the isotropic cluster (\textit{blue}) and observed stars (\textit{magenta}). The vertical lines in this case indicate high $|h|$ values corresponding to tangential orbits (\textit{yellow}), low $|h|$ values corresponding to radial orbits (\textit{orange}), and unbound orbits (\textit{red}). This indicates that radial orbits (or low $|h|$ values) are more common within the sample than orbits within the simulated isotropic cluster. Again, to best model observations, all members of the isotropic cluster with separations of $< 2\,$arcseconds from Sgr~A* are discarded before plotting. }
    \label{fig:abshbothnounbound}
\end{figure}

This inconsistency with the simulated isotropic cluster is also demonstrated in Fig. \ref{fig:abshbothnounbound}. The vertical lines in Fig. \ref{fig:abshbothnounbound} indicate low $|h|$ values (orange) corresponding to radial orbits, high $|h|$ values (yellow) indicating tangential orbits and unbound orbits (red) at $|h| = \sqrt{2}$. None of the stars appear above the $|h| = \sqrt{2}$ line due to the condition applied when searching for significant accelerations, which excludes all nonphysical (i.e., unbound) accelerations and the fact that the cluster members are bound to Sgr~A* by definition. Here it is clear that the observed stars have a different distribution of $|h|$ values than the isotropic cluster, shifted toward lower $|h|$ values. Comparison to the isotropic cluster also indicates that this sample may be biased to high eccentricity orbits since a higher proportion of measured stars feature $|h|\sim0$. Again, using the KS test now to compare the $|h|$ values for the simulated cluster and the observed stars, a KS score of $\sim 0.32$ is returned corresponding to a p value of $\sim 2.0 \times 10^{-15}$. This again indicates that we should reject the null hypothesis in this case and conclude the observed stars do not originate from the same isotropic distribution as the simulated cluster. Overall, the results suggest that the observed sample of stars is not well represented by an isotropic cluster.

In summary, these comparisons to either known disk features in the GC environment using Hammer-Aitoff projection maps or the simulated isotropic cluster using the $j$ and $h$ statistics are consistent with results of previous studies that have found that the GC environment is round but not isotropic due to disks and features being present in the stellar population. Further, disk membership could be a useful tool to lift orbital degeneracies and constrain the mass enclosed within the orbits of stars with full orbital solutions. However, further work is required to increase the precision of the radial velocity measurements in order to produce meaningful constraints on the enclosed mass in the GC.  
\section{Summary} \label{sec:summary}
This work extends the long-term monitoring program of stars in the GC to those located at greater separations from Sgr~A*. In order to accomplish this, we created a source finder that can track more than 1800 stars over 17 years of NACO data within the complex environment represented by the GC nuclear star cluster. We then calculated the proper motions for these stars using position information across these epochs in order to determine second-order fits for the motion of these stars in rotated radial and tangential components of motion. A subset that includes 54 of these stars presented significant astrometric accelerations in the direction of Sgr A*. For these stars, we used the measured 2D accelerations to determine the absolute value of the $z$ coordinate for each source.

Further, we determined radial velocities from SINFONI data for 20 of the stars with significant astrometric accelerations. The radial velocities could then be used to transform the six phase-space coordinates into orbital elements that describe the motion of the stars around Sgr~A*. This included three previously studied reference stars, as well as 17 stars that previously had no determined orbits. In total, we determined that ten of the stars are spectroscopically old and the remaining ten young. Three of these young stars had no previous spectroscopic classification. 

We also considered the orbital solutions of stars without radial velocity measurements using Monte Carlo sampling. By plotting the sets of physically possible solutions on Hammer-Aitoff projection maps, we identified trends in the orbits. Almost all of the stars were found to include ranges of orbital solutions that could classify them as disk members. If only one sign for the $z$ coordinate of these stars resulted in solutions that were consistent with a disk feature, the corresponding sign for the $z$ position was selected and the resulting orbit was used to further understand the GC environment. This was the case for 20 5D constrained stars. Of the stars with orbital solutions, three had values of inclination and longitude of the ascending node that corresponded to one of three disk features, namely CW1, CW2, or F2 (although the last case is unlikely due to separation considerations).
We then assumed the disk membership to be true in order to derive a $z$ coordinate that depends only on the $i$ and $\Omega$ of the disk to which the star belongs and the stellar positions in RA and Dec. We then derived the mass enclosed within the orbit of the star. We find that the median enclosed mass we calculated, when compared to previous works, agrees within uncertainties with the expected mass enclosed in the clockwise disk \citep{GRAVITYmassdist2021}. However, this measurement is not able to constrain the enclosed mass. This is likely due to the fact that the radial velocities for these stars are too uncertain.
The GC environment was also probed using the $j$ and $h$ statistics of the stars with significant radial accelerations toward Sgr~A* by comparing them to those calculated for members of a simulated isotropic cluster. We find that the sample is not well represented by the modeled isotropic cluster, which is consistent with the fact that disks and features are observed in the GC (\citealt{2disksPaumard2006,Bartko2010,Genzel_2010_gcreview}; vF22).
Future observations could be used to further this work, particularly in the field of IR spectroscopy. Within the next ten years, for example, it should be possible to measure a change in radial velocity greater than $50\,$km/s for the stars contained in Table \ref{tab:elapsetimeto50kmchange}, namely: 3-19, 4-120, 3-20, 4-56, 4-36, 3-53, 3-91, 5-246, 3-130, 3-161, 3-158, and 3-134. This would make it possible to constrain the $z$ coordinate for these stars, removing the degeneracy of their orbital solutions as well as allowing them to contribute to constraining the potential of Sgr~A*.

One instrument that will be crucial to studies of the GC in the coming years is ERIS, the installation of which was completed at the UT4 of the ESO VLT in early 2022 and which will serve as the replacement for NACO and SINFONI \citep{2018eris}. ERIS contains an upgraded version of SPIFFI, namely SPIFFIER, which should result in better throughput and higher spectral resolution. The NIR camera, NIX, is also expected to have a comparable performance to NACO. Specifically, ERIS is optimized for GC observing with the resolution and wavelength required for the continued monitoring of fainter and closer stars \citep{2018eris}.

Another exciting future instrument for the monitoring of stars featured in this work is the Multi-AO Imaging Camera for Deep Observations (MICADO), which will be the first instrument installed on the ESO Extremely Large Telescope \citep[ELT;][]{micado2021}. MICADO will have a resolution of only three times less than the current resolution of GRAVITY while also having a FOV comparable to that of NACO \citep{micado2021}. As such, MICADO will peer deeper into the outer regions of the GC than any previous instrument and will be essential to furthering this work. Beyond imaging, it is necessary to acquire spectra for stars that feature significant accelerations toward Sgr A*. Definite orbital solutions could still not be produced for 34 of the 5D constrained stars detected in this work, due to a lack of radial velocity measurements from stellar spectra. As such, spectroscopy from ERIS and, further into the future, MICADO and HARMONI on the ELT promises to provide improved resolution compared to SINFONI, which will accurately measure radial velocities for these stars. This increased precision will also contribute to improving the constraints on the enclosed mass within the orbits of stars at these separations. 
 
In conclusion, this work not only increases the number of known orbits within the GC by $\sim40\,\%$, it also confirms previous studies that have found evidence for multiple disk structures within the nuclear star cluster environment. In addition, this work provides strong scientific justification for follow-up observations using future instrumentation that will continue to provide valuable information about the stars at greater separations from Sgr~A*.
\begin{acknowledgements}
This publication is based on observations collected at the European Southern Observatory and the authors thank ESO and the ESO/Paranal staff, without whom this work would not be possible. We also thank the referee for providing useful comments which helped to improve the paper. AY acknowledges support from the European Space Agency's Young Graduate Trainee program. JS was supported in part by the Deutsche Forschungsgemeinschaft (DFG, German Research Foundation) under Germany's Excellence Strategy - EXC-2094 - 390783311.
\end{acknowledgements}


\bibliography{updated}
\bibliographystyle{aa} 


\onecolumn
\begin{appendix}
\section{List of NACO data sets} \label{AP:nacoappendix}
\begin{table}[htpb]
    \centering
    \caption{NACO data sets for observation epochs from 2002 to 2011.}
    \begin{tabular}{cccccccc} \hline\hline
    Epoch & Band & $\frac{mas}{pixel}$ & DIT [s] & NDIT & No. Frames & No. Stars Detected \\ \hline
20020731 & Ks & 13.27 & 15 & 4 & 60 & 888 \\
20030320 & H & 13.27 & 20 & 1 & 32 & 868 \\
20030905 & H & 13.27 & 20 & 3 & 32 & 862 \\
20040328 & H & 13.27 & 10 & 3 & 32 & 785 \\
20050513 & Ks & 13.27 & 2 & 15 & 24 & 912 \\
20050516 & Ks & 13.27 & 20 & 10 & 8 & 926 \\
20050727 & Ks & 13.27 & 15 & 4 & 30 & 911 \\
20050729 & Ks & 13.27 & 15 & 4 & 30 & 880 \\
20060429 & H & 13.27 & 17 & 2 & 24 & 900 \\
20060530 & H & 13.27 & 17 & 2 & 24 & 870 \\
20070304 & Ks & 13.27 & 12 & 3 & 24 & 636 \\
20070316 & Ks & 13.27 & 2 & 14 & 24 & 674 \\
20070319 & Ks & 13.27 & 17 & 2 & 24 & 680 \\
20070402 & Ks & 13.27 & 10 & 3 & 24 & 688 \\
20070618 & Ks & 13.27 & 17 & 2 & 24 & 998 \\
20070718 & H & 13.27 & 10 & 3 & 24 & 714 \\
20070720 & H & 13.27 & 10 & 3 & 24 & 700 \\
20070908 & H & 13.27 & 17 & 2 & 24 & 705 \\
20070909 & Ks & 13.27 & 17 & 2 & 24 & 704 \\
20070910 & Ks & 13.27 & 17 & 2 & 24 & 712 \\
20080313 & Ks & 13.27 & 17 & 2 & 24 & 1100 \\
20080616 & Ks & 13.27 & 17 & 2 & 24 & 1118 \\
20090515 & Ks & 13.27 & 3 & 12 & 24 & 1129 \\
20090722 & Ks & 13.27 & 18 & 2 & 6 & 1167 \\
20110330 & Ks & 13.27 & 15.5 & 20 & 4 & 1270 \\
20110330 & Ks & 13.27 & 15 & 20 & 4 & 1270 \\
20110424 & Ks & 13.27 & 15 & 20 & 4 & 1223 \\
20110503 & Ks & 13.27 & 4 & 67 & 4 & 1212 \\
20110611 & Ks & 13.27 & 4 & 67 & 4 & 1199 \\
20110721 & Ks & 13.27 & 4 & 67 & 4 & 1134 \\
20110812 & Ks & 13.27 & 4 & 67 & 4 & 1172 \\
20110909 & Ks & 13.27 & 4 & 67 & 4 & 1169 \\
20110911 & Ks & 13.27 & 4 & 67 & 4 & 1183 \\
20110912 & Ks & 13.27 & 4 & 67 & 4 & 1199 \\
20110921 & Ks & 13.27 & 4 & 67 & 4 & 1185 \\\hline
      \end{tabular}
      \tablefoot{DIT stands for single-detector integration time and NDIT for the number of single integrations per image file.}
    \label{tab:NACOdata1}
\end{table}
\newpage
\begin{table}[H]
    \centering
    \caption{Continued list of NACO data sets for observation epochs from 2012 to 2019. }
    \begin{tabular}{cccccccc} \hline\hline
    Epoch & Band & $\frac{mas}{pixel}$ & DIT [s] & NDIT & No. Frames & No. Stars Detected \\  \hline
    20120630 & H & 13.27 & 10 & 8 & 12 & 1158 \\
20120713 & Ks & 13.27 & 4 & 67 & 4 & 1187 \\
20120717 & Ks & 13.27 & 4 & 67 & 4 & 1188 \\
20120808 & Ks & 13.27 & 4 & 67 & 4 & 1200 \\
20130227 & Ks & 13.27 & 4 & 18 & 12 & 1200 \\
20130328 & Ks & 13.27 & 4 & 67 & 4 & 1224 \\
20130425 & Ks & 13.27 & 4 & 18 & 12 & 1176 \\
20130513 & Ks & 13.27 & 4 & 18 & 12 & 1213 \\
20130608 & Ks & 13.27 & 4 & 18 & 12 & 1160 \\
20130629 & Ks & 13.27 & 4 & 18 & 12 & 1184 \\
20130702 & Ks & 13.27 & 4 & 18 & 12 & 1179 \\
20130803 & Ks & 13.27 & 4 & 18 & 12 & 1092 \\
20130813 & Ks & 13.27 & 4 & 18 & 12 & 1142 \\
20150707 & Ks & 13.27 & 4 & 18 & 60 & 1034 \\
20150915 & Ks & 13.27 & 4 & 18 & 45 & 980 \\
20160428 & Ks & 13.27 & 4 & 18 & 16 & 991 \\
20160514 & Ks & 13.27 & 4 & 18 & 16 & 927 \\
20160713 & Ks & 13.27 & 4 & 18 & 23 & 949 \\
20170427 & H & 13.27 & 10 & 8 & 24 & 828 \\
20170430 & Ks & 13.27 & 4 & 18 & 24 & 913 \\
20170601 & Ks & 13.27 & 4 & 18 & 24 & 948 \\
20170615 & Ks & 13.27 & 4 & 18 & 36 & 983 \\
20170816 & Ks & 13.27 & 4 & 18 & 24 & 936 \\
20170821 & Ks & 13.27 & 4 & 18 & 24 & 965 \\
20170823 & Ks & 13.27 & 4 & 18 & 24 & 972 \\
20180321 & Ks & 13.27 & 4 & 18 & 48 & 1001 \\
20180326 & Ks & 13.27 & 4 & 18 & 24 & 990 \\
20180511 & Ks & 13.27 & 4 & 18 & 95 & 967 \\
20180517 & Ks & 13.27 & 4 & 18 & 42 & 1007 \\
20180524 & Ks & 13.27 & 4 & 18 & 24 & 974 \\
20180527 & Ks & 13.27 & 4 & 18 & 24 & 976 \\
20180608 & Ks & 13.27 & 4 & 18 & 51 & 941 \\
20180609 & Ks & 13.27 & 4 & 30 & 18 & 766 \\
20180612 & Ks & 13.27 & 4 & 18 & 29 & 898 \\
20180712 & Ks & 13.27 & 4 & 30 & 18 & 904 \\
20180713 & Ks & 13.27 & 4 & 30 & 18 & 877 \\
20180717 & H & 13.27 & 10 & 8 & 33 & 871 \\
20180808 & Ks & 13.27 & 4 & 18 & 48 & 944 \\
20180809 & H & 13.27 & 10 & 8 & 45 & 942 \\
20180817 & Ks & 13.27 & 4 & 18 & 24 & 909 \\
20190602 & Ks & 13.27 & 4 & 18 & 42 & 915 \\ \hline
      \end{tabular}
      \tablefoot{DIT stands for single-detector integration time and NDIT for the number of single integrations per image file.}
    \label{tab:NACOdata2}
\end{table}
\newpage
\section{List of SINFONI data sets} \label{AP:sinfappendix}
\begin{table}[H]
\caption{Information about the datacubes used to extract radial velocity information for stars with significant astrometric accelerations. }
    \centering
    \begin{tabular}{cccccccccccc} \hline\hline
    Epoch & Band & DIT [s] & NDIT & Epoch (con't) & Band (con't) & DIT [s]  (con't)& NDIT (con't) \\  \hline
    20040818-19 & K & 600 & 1 & 20140929 &      H+K & 300 &     1\\
    20110703 &  H+K & 100 &     6 & 20140929 &  H+K     & 300 & 1 \\
    20110703 &  H+K     & 100 & 6 & 20140929 &  H+K     & 300 & 1\\
    20140302 &  H+K     & 200 & 1 & 20140929 &  H+K     & 300 & 1 \\
    20140408 &  H+K     & 300 & 1 & 20140930 &  H+K     & 300 & 1 \\
    20140408 &  H+K     & 300 & 1 & 20160412 &  H+K     & 300 & 1 \\
    20140409 &  H+K     & 300 & 1 & 20160412 &  H+K     & 300 & 1 \\ 
    20140409 &  H+K     & 300 & 1 & 20160412 &  H+K     & 300 & 1 \\
    20140928 &  H+K     & 300 & 1 & 20160412 &  H+K     & 300 & 1\\
    20140928 &  H+K     & 300 & 1 & 20160412 &  H+K     & 300 & 1\\
    20140928 &  H+K     & 300 & 1 & &&&\\

      \hline
      \end{tabular}
      \tablefoot{The cubes listed include the August 2004 cube, which features complete atmosphere subtraction as well as the data used to construct additional cubes when spectroscopic lines were not detectable in the August 2004 cube. DIT: single-detector integration time, NDIT: number of single integrations per image file.}
    \label{tab:SINFONIdata}
\end{table}
\section{Polynomial fits for RA and Dec. motion of stars with significant astrometric acceleration} \label{AP:sigstarspolyfits}
\begin{table}[H]
    \centering
    \caption{Second-order polynomial coefficient fit results for all stars found to have significant astrometric accelerations within the inner 3 arcseconds surrounding Sgr~A* in Sect. \ref{sec:accelerations}.}
    \setlength{\tabcolsep}{2pt}
    \begin{tabular}{ccccccccccccc}\hline \hline
    ID & R.A.& $\sigma$\,R.A. & Dec.& $\sigma\,$Dec.& $v_{R.A.}$& $\sigma\,v_{R.A.}$& $v_{Dec.}$& $\sigma\,v_{Dec.}$& $a_{RA}$& $\sigma\,a{RA}$& $a_{Dec.}$& $\sigma\,a_{Dec.}$ \\
    &[mas] & [mas] & [mas] & [mas]& [mas/yr] & [mas/yr] &[mas/yr] &[mas/yr] & [mas/yr$^2$]&[mas/yr$^2$]&[mas/yr$^2$]&[mas/yr$^2$]\\\hline 
   3-3 & 1.25e-1 & 1.16e-3 & -2.88  & 1.36e-3 & 2.39e-3 & 1.95e-4 & 2.37e-3 & 1.92e-4 & -1.53e-4 & 4.17e-5 & 3.15e-4 & 4.13e-5 \\
3-4 & -7.16e-1 & 5.32e-4 & -2.87  & 2.36e-4 & 3.02e-3 & 2.61e-4 & -4.19e-4 & 3.53e-5 & 1.90e-5 & 2.62e-5 & 6.49e-5 & 5.42e-6 \\
3-7 & -4.87e-1 & 1.52e-4 & -2.79  & 2.43e-4 & -1.02e-3 & 5.05e-5 & -1.29e-3 & 8.19e-5 & 7.67e-5 & 8.40e-6 & 1.99e-6 & 1.58e-5 \\
3-9 & 9.86e-1 & 2.28e-4 & -2.78  & 2.26e-4 & -2.60e-3 & 5.23e-5 & 8.53e-4 & 5.12e-5 & 3.14e-5 & 1.25e-5 & 6.32e-5 & 1.23e-5 \\
3-17 & -8.79e-1 & 5.91e-5 & -2.53  & 6.94e-5 & 4.49e-3 & 1.18e-5 & -8.81e-4 & 1.47e-5 & 4.68e-6 & 2.24e-6 & 8.37e-6 & 2.83e-6 \\
3-18 & 1.26  & 4.44e-4 & -2.54  & 4.24e-4 & -4.97e-3 & 7.88e-5 & -1.54e-3 & 9.45e-5 & -6.06e-5 & 1.36e-5 & 2.50e-5 & 1.47e-5 \\
3-19 & 4.11e-1 & 1.11e-4 & -2.48  & 8.28e-5 & 1.06e-3 & 3.59e-5 & 1.48e-3 & 2.65e-5 & -4.73e-5 & 5.89e-6 & 2.24e-5 & 4.49e-6 \\
3-20 & 1.66  & 6.99e-5 & -2.47  & 7.65e-5 & 6.11e-4 & 1.72e-5 & -1.69e-3 & 1.97e-5 & 1.13e-5 & 3.40e-6 & 2.42e-5 & 3.97e-6 \\
3-27 & -1.88  & 9.92e-5 & -2.34  & 1.14e-4 & -2.81e-3 & 2.35e-5 & 1.27e-2 & 2.78e-5 & -2.49e-5 & 4.12e-6 & 3.28e-5 & 4.98e-6 \\
3-40 & 9.03e-1 & 6.92e-5 & -2.19  & 8.60e-5 & 3.66e-3 & 1.55e-5 & 3.15e-4 & 1.90e-5 & 2.04e-5 & 3.91e-6 & 3.17e-5 & 5.17e-6 \\
3-47 & 1.95  & 6.13e-5 & -2.16  & 1.18e-4 & 1.08e-3 & 1.43e-5 & -2.10e-3 & 2.92e-5 & 4.80e-6 & 2.78e-6 & 2.47e-5 & 5.85e-6 \\
3-52 & -4.72e-1 & 1.05e-4 & -2.09  & 1.21e-4 & 1.59e-4 & 1.67e-5 & 1.13e-3 & 2.38e-5 & 3.14e-5 & 3.03e-6 & 3.05e-5 & 4.01e-6 \\
3-53 & -1.15  & 4.06e-5 & -2.09  & 5.54e-5 & 8.68e-3 & 1.19e-5 & 3.89e-3 & 1.64e-5 & -1.53e-5 & 3.00e-6 & 3.16e-5 & 4.27e-6 \\
3-58 & 1.98  & 9.12e-5 & -2.04  & 9.47e-5 & -5.00e-4 & 3.34e-5 & 3.55e-3 & 3.27e-5 & -5.29e-5 & 6.06e-6 & 2.55e-5 & 6.06e-6 \\
3-72 & -2.43  & 1.61e-4 & -1.69  & 1.20e-4 & 4.92e-4 & 4.14e-5 & 2.34e-3 & 3.29e-5 & -6.04e-6 & 6.92e-6 & 3.69e-5 & 5.45e-6 \\
3-74 & -1.62  & 6.27e-5 & -1.68  & 5.07e-5 & 9.49e-3 & 1.32e-5 & -3.88e-3 & 1.14e-5 & 9.18e-6 & 2.07e-6 & 2.69e-5 & 1.72e-6 \\
3-84 & 1.52  & 6.12e-5 & -1.46  & 6.53e-5 & 8.23e-3 & 1.15e-5 & 2.73e-3 & 1.22e-5 & -3.10e-5 & 2.27e-6 & 2.78e-6 & 2.39e-6 \\
3-87 & 2.31  & 1.49e-4 & -1.41  & 1.71e-4 & 2.32e-3 & 2.64e-5 & -1.59e-3 & 3.14e-5 & -2.03e-5 & 5.12e-6 & 7.22e-5 & 6.33e-6 \\
3-89 & -2.51  & 6.96e-4 & -1.35  & 4.64e-4 & 1.41e-3 & 1.41e-4 & 2.54e-3 & 9.36e-5 & 5.86e-5 & 2.28e-5 & 9.82e-5 & 1.54e-5 \\
3-91 & 1.66  & 7.48e-5 & -1.34  & 8.16e-5 & 8.07e-3 & 1.21e-5 & 2.02e-3 & 1.34e-5 & -1.05e-5 & 2.56e-6 & 2.98e-5 & 2.89e-6 \\
3-103 & 1.90  & 4.98e-5 & -1.12  & 4.09e-5 & 5.99e-3 & 1.12e-5 & 3.72e-3 & 8.75e-6 & -2.16e-5 & 2.10e-6 & 1.15e-5 & 1.72e-6 \\
3-104 & -2.72  & 2.00e-4 & -1.05  & 4.50e-4 & 9.60e-3 & 5.21e-5 & -6.96e-3 & 1.25e-4 & 7.40e-5 & 7.83e-6 & -1.38e-4 & 1.84e-5 \\
3-105 & 2.37  & 8.81e-5 & -1.08  & 1.12e-4 & 6.19e-4 & 2.23e-5 & -2.83e-3 & 2.69e-5 & 4.02e-6 & 5.09e-6 & 8.51e-5 & 7.81e-6 \\
3-111 & 2.64  & 6.63e-5 & -8.54e-1 & 7.99e-5 & -2.58e-4 & 1.44e-5 & 4.63e-3 & 1.76e-5 & -3.45e-5 & 2.67e-6 & 6.19e-5 & 3.22e-6 \\
3-114 & 1.94  & 6.80e-5 & -7.88e-1 & 1.18e-4 & 5.62e-3 & 1.51e-5 & 5.07e-3 & 2.78e-5 & -3.66e-5 & 4.13e-6 & 7.68e-6 & 7.74e-6 \\
3-134 & 2.30  & 6.87e-5 & -2.07e-1 & 8.89e-5 & -1.79e-3 & 1.32e-5 & 6.18e-3 & 1.71e-5 & -3.74e-5 & 2.36e-6 & 2.08e-5 & 3.11e-6 \\
3-142 & -2.80  & 9.95e-4 & 1.40e-1 & 9.27e-4 & -3.22e-3 & 2.21e-4 & -3.00e-3 & 2.10e-4 & -6.76e-5 & 4.48e-5 & 3.45e-4 & 4.22e-5 \\
3-158 & 1.99  & 8.49e-5 & 4.44e-1 & 2.16e-4 & -5.07e-3 & 2.18e-5 & 5.16e-3 & 5.20e-5 & -5.24e-5 & 3.97e-6 & 1.96e-5 & 8.97e-6 \\
3-161 & 1.99  & 7.84e-5 & 5.97e-1 & 5.49e-5 & -2.01e-3 & 3.15e-5 & 6.22e-3 & 1.98e-5 & -2.97e-5 & 4.40e-6 & 2.85e-6 & 3.06e-6 \\
3-190 & -2.38  & 5.04e-4 & 1.49  & 7.69e-4 & -4.38e-3 & 9.68e-5 & 4.60e-3 & 1.18e-4 & 3.51e-5 & 1.83e-5 & -1.07e-4 & 2.36e-5 \\
3-244 & 7.70e-1 & 9.40e-5 & 2.74  & 1.96e-4 & 1.28e-2 & 3.26e-5 & 1.55e-4 & 7.03e-5 & 2.59e-5 & 4.67e-6 & -7.60e-5 & 1.06e-5 \\
3-81 & -1.33  & 8.60e-4 & -1.50  & 4.84e-4 & -5.70e-3 & 1.40e-4 & 5.28e-3 & 1.66e-4 & 2.03e-4 & 2.94e-5 & 1.70e-6 & 2.90e-5 \\
3-143 & 2.74  & 3.22e-3 & 3.24e-1 & 7.26e-2 & 1.22e-3 & 7.56e-4 & -9.69e-2 & 7.47e-3 & 3.02e-3 & 1.10e-4 & -4.17e-2 & 1.65e-3 \\
    \hline
   \end{tabular}
    \label{tab:significantstarspolyfitsinner}
\end{table}
\newpage
\begin{table}[H]
    \centering
     \caption{Second-order polynomial coefficient fit results for all stars found to have significant astrometric accelerations outside the inner 3 arcseconds surrounding Sgr~A* in Sect. \ref{sec:accelerations}.}
    \setlength{\tabcolsep}{2pt}
    \begin{tabular}{ccccccccccccc}\hline \hline
     ID & R.A.& $\sigma$\,R.A. & Dec.& $\sigma\,$Dec.& $v_{R.A.}$& $\sigma\,v_{R.A.}$& $v_{Dec.}$& $\sigma\,v_{Dec.}$& $a_{RA}$& $\sigma\,a_{RA}$& $a_{Dec.}$& $\sigma\,a_{Dec.}$ \\
    &[mas] & [mas] & [mas] & [mas]& [mas/yr] & [mas/yr] &[mas/yr] &[mas/yr] & [mas/yr$^2$]&[mas/yr$^2$]&[mas/yr$^2$]&[mas/yr$^2$]\\\hline 
    4-19 & -9.80e-1 & 7.45e-5 & -3.42  & 9.63e-5 & 6.19e-3 & 1.24e-5 & -2.95e-3 & 1.58e-5 & 8.45e-6 & 2.36e-6 & 1.02e-5 & 3.01e-6 \\
4-29 & 1.28  & 1.12e-4 & -3.37  & 9.34e-5 & 2.07e-3 & 2.34e-5 & -4.68e-4 & 2.38e-5 & 7.74e-6 & 3.85e-6 & 1.33e-5 & 3.85e-6 \\
4-36 & 1.33  & 3.53e-4 & -3.31  & 1.56e-4 & 2.59e-3 & 1.08e-4 & 3.63e-5 & 5.04e-5 & -1.44e-4 & 1.67e-5 & -1.11e-5 & 7.76e-6 \\
4-46 & -1.85  & 1.61e-4 & -3.18  & 1.90e-4 & -1.53e-3 & 4.02e-5 & -4.27e-3 & 4.84e-5 & -2.93e-6 & 7.19e-6 & 4.16e-5 & 9.03e-6 \\
4-56 & 1.44  & 6.74e-5 & -3.08  & 6.56e-5 & 2.23e-3 & 1.29e-5 & 1.63e-3 & 1.34e-5 & -1.18e-5 & 2.65e-6 & 2.13e-5 & 2.87e-6 \\
4-62 & -1.39  & 5.32e-5 & -3.02  & 5.52e-5 & 1.62e-3 & 1.33e-5 & 1.53e-3 & 1.41e-5 & -7.91e-6 & 2.79e-6 & 2.10e-5 & 3.14e-6 \\
4-64 & -8.49e-1 & 4.52e-5 & -3.00  & 6.85e-5 & 1.46e-3 & 9.76e-6 & 3.07e-3 & 1.63e-5 & 4.64e-6 & 1.98e-6 & 2.86e-5 & 3.30e-6 \\
4-79 & -1.55  & 9.57e-5 & -2.80  & 5.81e-5 & 7.94e-3 & 1.56e-5 & -1.85e-3 & 9.61e-6 & 1.11e-5 & 3.29e-6 & 1.11e-5 & 2.03e-6 \\
4-83 & 1.58  & 7.28e-5 & -2.79  & 6.65e-5 & -1.94e-4 & 2.13e-5 & 6.76e-4 & 1.92e-5 & -4.99e-6 & 4.95e-6 & 2.28e-5 & 5.35e-6 \\
4-109 & -2.94  & 4.50e-4 & -1.98  & 2.55e-4 & 3.07e-3 & 1.11e-4 & -1.57e-3 & 6.19e-5 & 5.95e-5 & 3.70e-5 & 1.16e-4 & 2.00e-5 \\
4-120 & 2.96  & 8.15e-5 & -1.91  & 1.16e-4 & -7.31e-5 & 1.71e-5 & 2.68e-3 & 2.50e-5 & -1.74e-5 & 3.36e-6 & 2.49e-5 & 5.10e-6 \\
4-207 & 3.84  & 3.67e-4 & 4.73e-1 & 6.67e-4 & 4.87e-4 & 7.06e-5 & -2.15e-3 & 1.17e-4 & -3.88e-5 & 1.46e-5 & 1.27e-4 & 1.96e-5 \\
4-241 & -3.07  & 1.03e-4 & 1.41  & 9.04e-5 & 2.72e-3 & 3.19e-5 & -2.14e-3 & 3.10e-5 & 3.75e-5 & 5.75e-6 & 1.58e-5 & 5.76e-6 \\
4-311 & -1.97  & 2.56e-4 & 3.41  & 2.52e-4 & -6.52e-3 & 4.47e-5 & 1.56e-3 & 4.33e-5 & 9.42e-6 & 9.41e-6 & -2.23e-5 & 9.99e-6 \\
4-326 & -9.61e-1 & 1.25e-4 & 3.80  & 1.04e-4 & -8.13e-3 & 2.17e-5 & -7.47e-4 & 1.91e-5 & 5.17e-6 & 4.33e-6 & -2.48e-5 & 3.76e-6 \\
4-329 & 6.29e-1 & 2.51e-4 & 3.90  & 1.50e-4 & -3.36e-3 & 4.29e-5 & -3.02e-3 & 2.67e-5 & 1.53e-5 & 9.47e-6 & -4.57e-5 & 5.77e-6 \\
4-272 & 2.75  & 7.48e-4 & 2.42  & 1.83e-3 & 1.17e-3 & 1.38e-4 & 4.13e-3 & 3.78e-4 & -5.34e-5 & 2.73e-5 & -1.89e-4 & 6.58e-5 \\
4-211 & -3.56  & 1.78e-4 & 6.61e-1 & 3.17e-4 & -3.19e-3 & 5.77e-5 & 3.10e-4 & 1.07e-4 & 4.64e-6 & 7.88e-6 & -6.32e-5 & 1.43e-5 \\
5-55 & 1.60  & 1.39e-4 & -3.77  & 1.47e-4 & -1.58e-3 & 3.75e-5 & -5.76e-4 & 3.89e-5 & -2.37e-5 & 6.91e-6 & 2.15e-5 & 7.31e-6 \\
5-139 & -4.15  & 1.16e-4 & -5.16e-3 & 1.71e-4 & -1.22e-3 & 4.41e-5 & -1.87e-3 & 6.06e-5 & 1.71e-5 & 5.97e-6 & -3.85e-5 & 8.19e-6 \\
5-246 & -9.96e-1 & 1.29e-4 & 3.99  & 7.78e-5 & 3.32e-3 & 2.63e-5 & -1.64e-3 & 1.59e-5 & 5.47e-5 & 5.65e-6 & -1.63e-5 & 3.70e-06 \\
    \hline
   \end{tabular}
    \label{tab:significantstarspolyfitsouter}
\end{table}

\section{Orbital elements for stars with measured radial velocities} \label{AP: orbitalel}
\begin{table}[H]
    \centering 
        \caption{Orbital elements that describe the motions of the 20 stars for which it was possible to determine all six phase-space coordinates.}
    \setlength{\tabcolsep}{2pt}
    \begin{tabular}{ccccccccccccccccccc}  \hline \hline
     ID & $a$  & $da$ & $e$ & $de$ & $i$ &  $di$  & $\Omega$ & $d\Omega$ & $\omega$  & $d\omega$  & $\theta$ & $d\theta$ & $t_p$  & $dt_p$ &  $P$ &  $dP$ & $v_{rad}$ & $dv_{rad}$  \\
     &[as] & [as] & & & $[^{\circ}]$ & $[^{\circ}]$ & $[^{\circ}]$ &$[^{\circ}]$ &$[^{\circ}]$ &$[^{\circ}]$ &$[^{\circ}]$ &$[^{\circ}]$ &[yr]&[yr]&[yr]&[yr]&[km/s]&[km/s] \\\hline
    3-19  &  1.33  &  0.0057  &  0.98  &  0.0058  &  143.8  &  9.58  &  329  &  15.0  &  334  &  13.5  &  181.2  &  0.25  &  2253  &  1.00  &  556.4  &  3.6  & -47.9 & 15\\
    3-20  &  1.64  &  0.0046 & 0.99 & 0.0047 & 43.5 & 15.75 & 352 & 139.0 & 327 & 44.4 & 179.4 & 0.2 & 1691 & 5.95 & 762.9 &  3.24 & 13.5 & 20\\
    3-53 & 2.79 & 0.19 & 0.84 & 0.01 & 150.7 & 1.1 & 306 & 13.0 & 242 & 9.8 & 214 & 1.53 & 2201.0 & 3.97 & 1690.3 & 183.9 & -181.3 & 30\\
    3-74 & 4.15 & 1.06 & 0.60 & 0.082 & 141.7 & 1.5 & 334 & 9.4 & 203 & 2.97 & 262 & 4.7 & 2269 & 22.4 & 3072 &  1336 & -210.2 & 53\\
    3-84 & 3.29 & 0.68 &  0.70 & 0.037 & 135.89 & 2.0 & 192 & 6.7 & 303 & 1.58 & 123 & 6.08 & 1787 & 10.8 & 2165 &  752.2 & 286 & 39\\
    3-89 & 2.83 & 1.6 &  0.38 & 0.22 & 80.4 & 12.5 & 235 & 45.6 & 251 & 40.1 & 145 & 42.4 & 1466 & 420 & 1733 & 1808 & 292.3 & 270\\
    3-91 & 2.32 & 0.11 & 0.73 & 0.012 & 141.4 & 0.49 & 212 & 6.8 & 299 & 3.4 & 145 & 1.96 & 1785 & 5.97 & 1286 &  93.9 & 182.5 & 25\\
    3-103 & 4.75 & 1.7 & 0.79 & 0.05 & 126.2 & 3.2 & 163 & 7.6 & 297 & 5.2 & 120 & 9.7 & 1779 & 14.2 & 3760 & 2217 & 366 & 70\\
    3-111 & 2.24 & 0.1 & 0.68 & 0.023 & 129.0 & 2.4 & 260 & 2.6 & 296 & 2.2 & 203 & 3.7 & 2345 & 0.68 & 1221 &  82.5 & -223.1 & 20\\
    3-134 & 1.57 & 0.177 & 0.62 & 0.13 & 144.7 & 9.45 & 134 & 27.7 & 218 & 21.3 & 186.3 & 12.3 & 1610 & 251.2 & 717 & 136.9 & 90 & 40 \\
    3-158 & 1.62 & 0.77 & 0.39 & 0.2 & 128.3 & 10.7 & 97 & 15.8 & 197 & 67.3 & 193 & 62.1 & 2329 & 151.5 & 749 &  789.6 & 224.1 & 127 \\
    3-161 & 2.18 & 0.16 & 0.50 & 0.011 & 127.7 & 1.84 & 116 & 3.53 & 272 & 5.92 & 144 & 8.03 & 1694 & 10.8 & 1170 & 131.5 & 241 & 30\\
    4-19 & 4.30 & 1.2 & 0.43 & 0.07 & 138.0 & 4.2 & 246 & 12.5 & 302 & 10.6 & 116 & 19.4 & 1421 & 96.8 & 3240 & 1522 & 185 & 65\\
    4-36 & 2.50 & 0.06 & 0.91 & 0.01 & 137.6 & 3.6 & 225 & 17.6 & 257 & 11.4 & 175.6 & 1.7 & 1436 & 26.9 & 1440 &  52.5 & 65 & 30 \\
    4-56 & 1.81 & 0.007 & 0.91 & 0.008 & 167.0 & 5.6 & 218 & 39.6 & 242 & 38.2 & 180.97 & 0.25 & 2433 & 6.9 & 888 & 5.0 & 7 & 25\\
    4-79 & 3.88 & 0.5 & 0.40 & 0.07 & 143.7 & 1.5 & 312 & 11.9 & 216 & 1.86 & 244.4 & 8.63 & 2541 & 64.5 & 2770 &  584.9 & -114 & 50 \\
    4-120 & 1.99 & 0.02 & 0.94 & 0.01 & 156.4 & 6.6 & 242 & 33.5 & 295 & 29.8 & 183.1 & 0.79 & 2427 & 8.24 & 1020 & 18.8 & -41.4 & 30\\
    4-207 & 6.44 & 1.6 & 0.66 & 0.1 & 75 & 11.5 & 70 & 111.6 & 286 & 22.9 & 115.8 & 28.8 & 1430 & 564.5 & 5900 &  2021 & 320.7 & 300\\
    4-326 & 4.03 & 0.42 & 0.19 & 0.077 & 158 & 3.05 & 69 & 25.9 & 329 & 87.4 & 114.8 & 31.8 & 1244 & 166.0 & 2940 & 514.6 & 35 & 70 \\
    5-246 & 2.89 & 0.07 & 0.82 & 0.07 & 45.7 & 6.95 & 300 & 20.1 & 229 & 13.1 & 186.9 & 2.07 & 2709 & 193.7 & 1789 & 71.0 & 10 & 50\\
 \hline
\end{tabular}
\tablefoot{Values listed in this table utilized the positive $z$ coordinate calculated from the 2D acceleration for each star.}
    \label{tab:orbitalelementsposz}
\end{table}
\newpage
\begin{table}
    \setlength{\tabcolsep}{2pt}
    \centering
     \caption{Orbital elements that describe the motions of the 20 stars for which it was possible to determine all six phase-space coordinates.}
    \begin{tabular}{ccccccccccccccccccc} \hline \hline
     ID & $a$  & $da$ & $e$ & $de$ & $i$ &  $di$  & $\Omega$ & $d\Omega$ & $\omega$  & $d\omega$  & $\theta$ & $d\theta$ & $t_p$  & $dt_p$ &  $P$ &  $dP$ & $v_{rad}$ & $dv_{rad}$ \\
     &[as] & [as] & & & $[^{\circ}]$ & $[^{\circ}]$ & $[^{\circ}]$ &$[^{\circ}]$ &$[^{\circ}]$ &$[^{\circ}]$ &$[^{\circ}]$ &$[^{\circ}]$&[yr]&[yr]&[yr]&[yr]&[km/s] &[km/s]  \\\hline
     3-19 &  1.328 & 0.0057 & 0.966 & 0.0098 & 129.1 & 6.9 & 3 & 109.6 & 18 & 2.8 & 181.0 & 0.053 & 2267 & 3.86 & 556.45 & 3.58 & -47.9 & 15\\
    3-20 &  1.639 & 0.0048 & 0.988 & 0.011 & 68.0 & 12.6 & 155.6 & 18.8 & 156.8 & 15.3 &  179.0 & 0.24 & 1682.3 & 7.9 & 762.9 & 3.39 & 13.5 & 20\\
    3-53 &  2.79 & 0.19 & 0.416 & 0.042 & 121.4 & 1.94 & 48.6 & 1.6 & 326.6 & 12.94 & 248.1 & 11.7 & 2308.4 & 9.6 & 1690.3 & 180.5 & -181.3 & 30\\
    3-74 &  4.15 & 1.09 & 0.301 & 0.13 & 131.3 & 2.57 & 84.9 & 4.76 & 229.5 & 24.6 & 3 & 163.2 & 1996.9 & 91.6 & 3072 & 1372 & -210.2 & 53\\
    3-84 &  3.29 & 0.7 & 0.21 & 0.13 & 122.0 & 1.84 & 103.2 & 2.47 & 357 & 159.5 & 314.6 & 28.7 & 2190.3 & 140.8 & 2165 & 770.4 & 286 & 39\\
    3-89 &  2.83 & 1.54 & 0.81 & 0.11 & 74.7 & 13.8 & 253.3 & 101.3 & 115.7 & 32.0 & 206.7 & 24.7 & 2316.5 & 228.3 & 1732.6 & 1771.3 & 292.3 & 270\\
    3-91 &  2.32 & 0.11 & 0.204 & 0.068 & 123.24 & 1.41 & 97.5 & 1.89 & 158.5 & 10.5 & 153.5 & 9.77 & 1505.4 & 66.7 & 1286.4 & 93.3 & 182.5 & 25\\
    3-103 &  4.75 & 1.73 & 0.48 & 0.15 & 114.6 & 2.2 & 95.5 & 2.9 & 15.3 & 22.8 & 296.5 & 24.1 & 2244.0 & 93.0 & 3757 & 2205 & 366 & 70\\
    3-111 &  2.24 & 0.099 & 0.46 & 0.046 & 121.4 & 1.7 & 308.4 & 1.5 & 49.1 & 4.1 & 166.6 & 4.8 & 1507.9 & 6.11 & 1220.6 & 81.1 & -223.1 & 20\\
    3-134 &  1.573 & 0.313 & 0.703 & 0.047 & 153.9 & 7.19 & 30.6 & 131.7 & 100.2 & 25.4 & 192.98 & 9.25 & 2268.1 & 31.4 & 717.3 & 524.5 & 90 & 40\\
    3-158 &  1.62 & 0.73 & 0.69 & 0.05 & 141.7 & 11.1 & 45.2 & 100.4 & 114.6 & 20.4 & 206.6 & 18.9 & 2197.0 & 10.8 & 749.137 & 726.0 & 224.1 & 127\\
    3-161 &  2.184 & 0.156 & 0.515 & 0.0098 & 128.28 & 1.82 & 29.87 & 3.6 & 87.2 & 5.26 & 216.03 & 7.44 & 2319.7 & 9.89 & 1173.1 & 127.8 & 241 & 30 \\
    4-19 &  4.298 & 1.2 & 0.208 & 0.09 & 133.28 & 4.4 & 155.8 & 9.0 & 51.7 & 32.3 & 257.4 & 38.4 & 2719.0 & 380.9 & 3240.0 & 1512.4 & 185 & 65 \\
    4-36 &  2.50 & 0.06 & 0.855 & 0.035 & 127.3 & 4.91 & 118.4 & 9.91 & 125.1 & 4.8 & 181.1 & 1.9 & 2703.9 & 590.4 & 1444.6 & 52.2  & 65 & 30\\
    4-56 &  1.814 & 0.007 & 0.915 & 0.007 & 168.4 & 5.10 & 72.1 & 58.8 & 95.82 & 40.0 & 181.11 & 0.29 & 2430.3 & 5.3 & 888.1 & 5.2 & 7 & 25 \\
    4-79 &  3.88 & 0.50 & 0.02 & 0.08 & 137.53 & 2.99 & 80.2 & 8.0 & 14.8 & 94.0 & 224.6 & 94.9 & 3041.6 & 590.7 & 2773.0 & 584.4 & -114 & 50\\
    4-120 &  1.989 & 0.024 & 0.909 & 0.029 & 140.4 & 9.2 & 330.0 & 84.9 & 31.6 & 14.3 & 182.2 & 0.26 & 2466.2 & 26.4 & 1020.3 & 18.8 & -41.4 & 30\\
    4-207 &  6.441 & 1.65 & 0.628 & 0.11 & 75.6 & 11.6 & 95.5 & 51.7 & 69.6 & 23.8 & 248.6 & 30.8 & 2609.9 & 592.3 & 5943.1 & 2070.5 & 320.7 & 300\\
    4-326 &  4.028 & 0.42 & 0.114 & 0.068 & 156.4 & 3.99 & 279.2 & 24.3 & 171.2 & 60.8 & 120.5 & 80.0 & 1122.9 & 320.3 & 2938.8 & 524.0 & 35 & 70\\    
    5-246 &  2.893 & 0.074 & 0.845 & 0.053 & 42.26 & 6.32 & 111.9 & 23.2 & 54.4 & 15.7 & 187.43 & 2.00 & 2677.4 & 120.5 & 1789.5 & 70.9 & 10 & 50 \\
 \hline
\end{tabular}
\tablefoot{Values listed in this table utilized the negative $z$ coordinate calculated from the 2D acceleration for each star.}
    \label{tab:orbitalelementsnegz}
\end{table}
\section{Coordinate system used for Hammer-Aitoff diagrams} \label{AP: othercoordhamaps}
\begin{figure}[h!]
  \centering
    {\includegraphics[width=.75\textwidth]{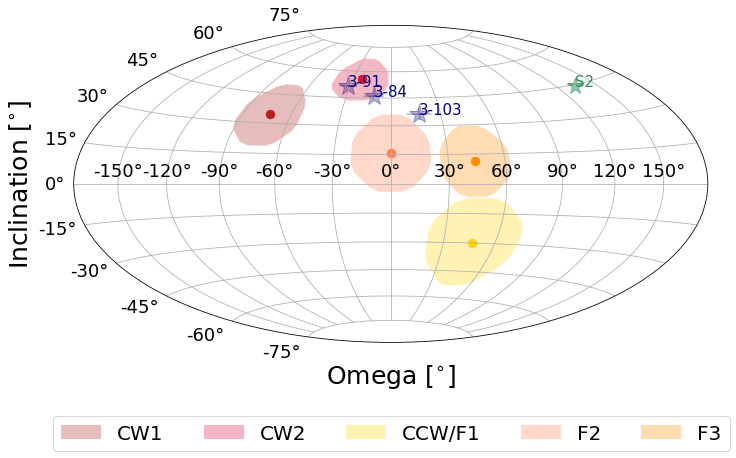}}
     \caption{Demonstration of the coordinate system used in previous works for Hammer-Aitoff projection maps (see, e.g., \cite{Gillessen2009,Gillessen2017}) by mapping Fig. \ref{fig:knownstarspoints} to this coordinate system. The transformation is most simply described as a 90 degree rotation  in $\Omega$. The projected ($\Omega$, i) coordinate corresponding to the orbital solution for S2 from \cite{Gillessen2017} is also included for reference (\textit{green}) in addition to the points representing the three stars also shown in Fig. \ref{fig:knownstarspoints} (\textit{navy}).}
  \label{fig:oldcoordinates}
\end{figure}

The coordinate system used to plot the Hammer-Aitoff projection maps in this work follows the conventions used in \cite{Sebyoungstars}. It relates to previous works that use different conventions for the plotted coordinates such as \cite{2disksPaumard2006}, \cite{Gillessen2009}, and \cite{Gillessen2017} by rotating $\Omega$ by 90 degrees relative to the coordinate system used in this work. To demonstrate this, Fig. \ref{fig:oldcoordinates} shows the equivalent of Fig. \ref{fig:knownstarspoints} in the coordinate system used in previous works. 
\newpage
\section{Distributions used to define the simulated isotropic cluster} \label{AP: isodist}
 \begin{figure}[H]
  \centering
        \label{subfig:semimaj}{%
        \begin{subfigure}[b]{0.45\textwidth}
            \centering
            \includegraphics[width=1\textwidth]{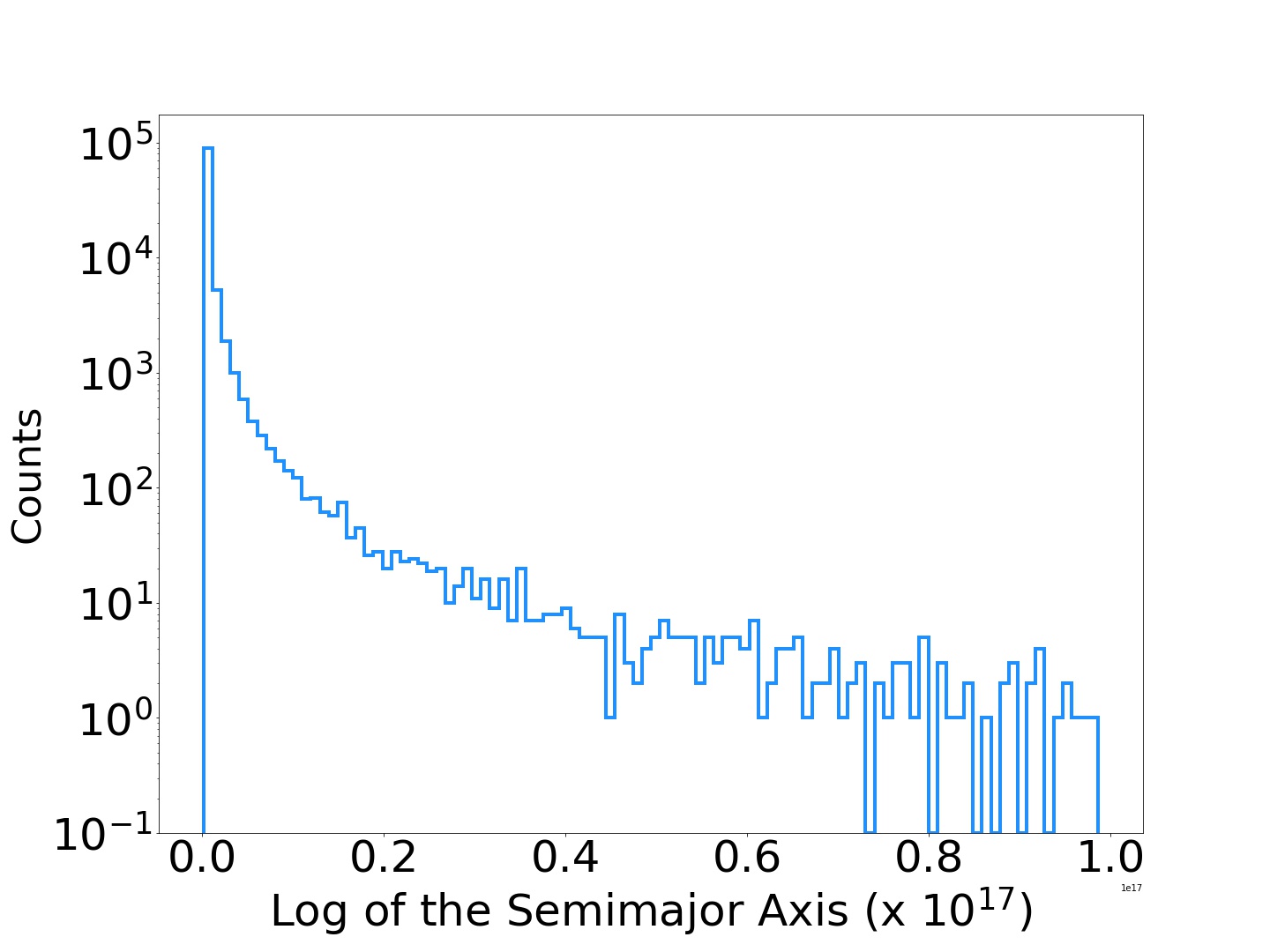}
        \end{subfigure}
        \hfill
        \begin{subfigure}[b]{0.45\textwidth}
            \centering
            \includegraphics[width=1\textwidth]{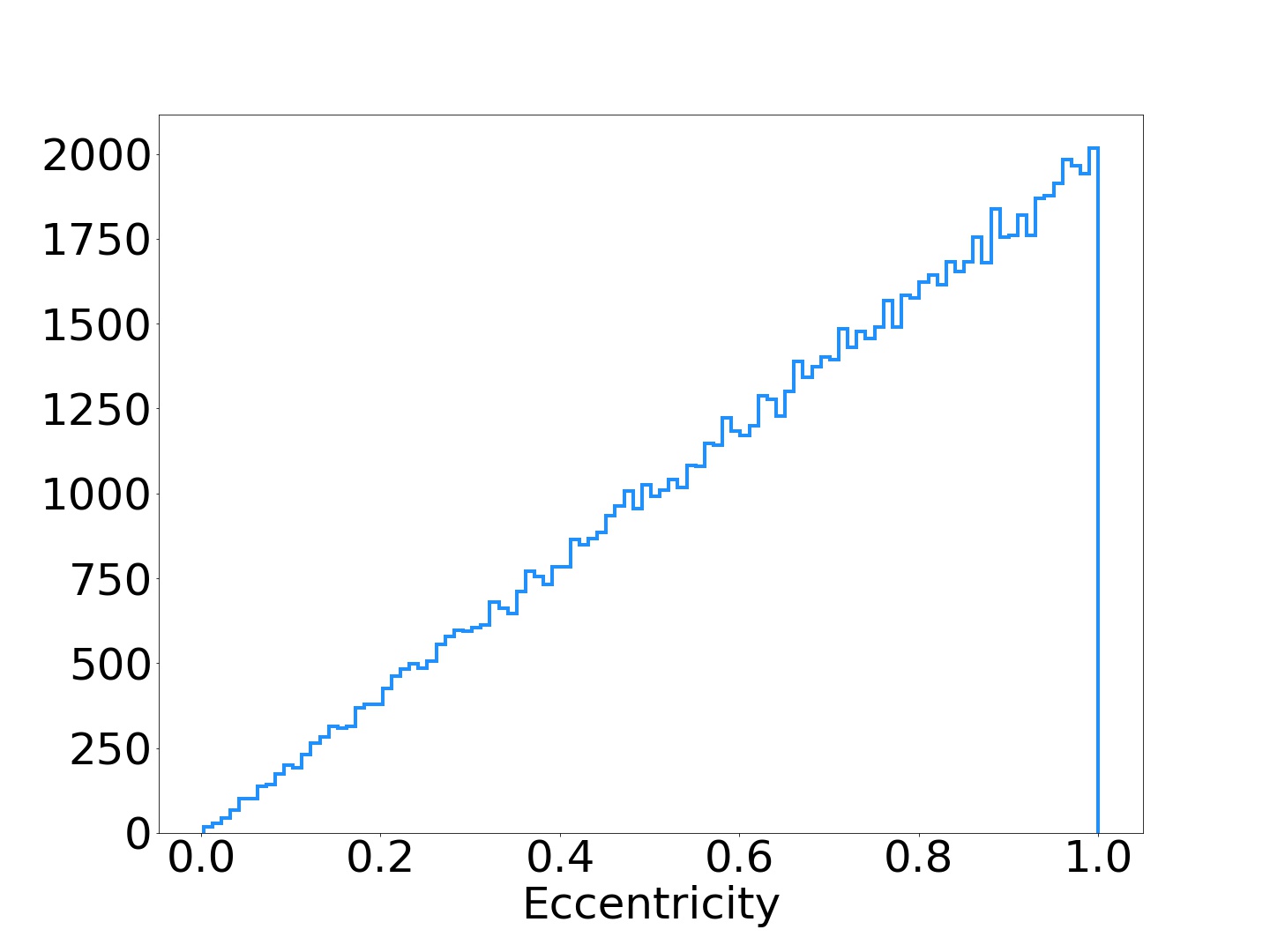}
        \end{subfigure}
        \hfill
        \begin{subfigure}[b]{0.45\textwidth}
            \centering
            \includegraphics[width=1\textwidth]{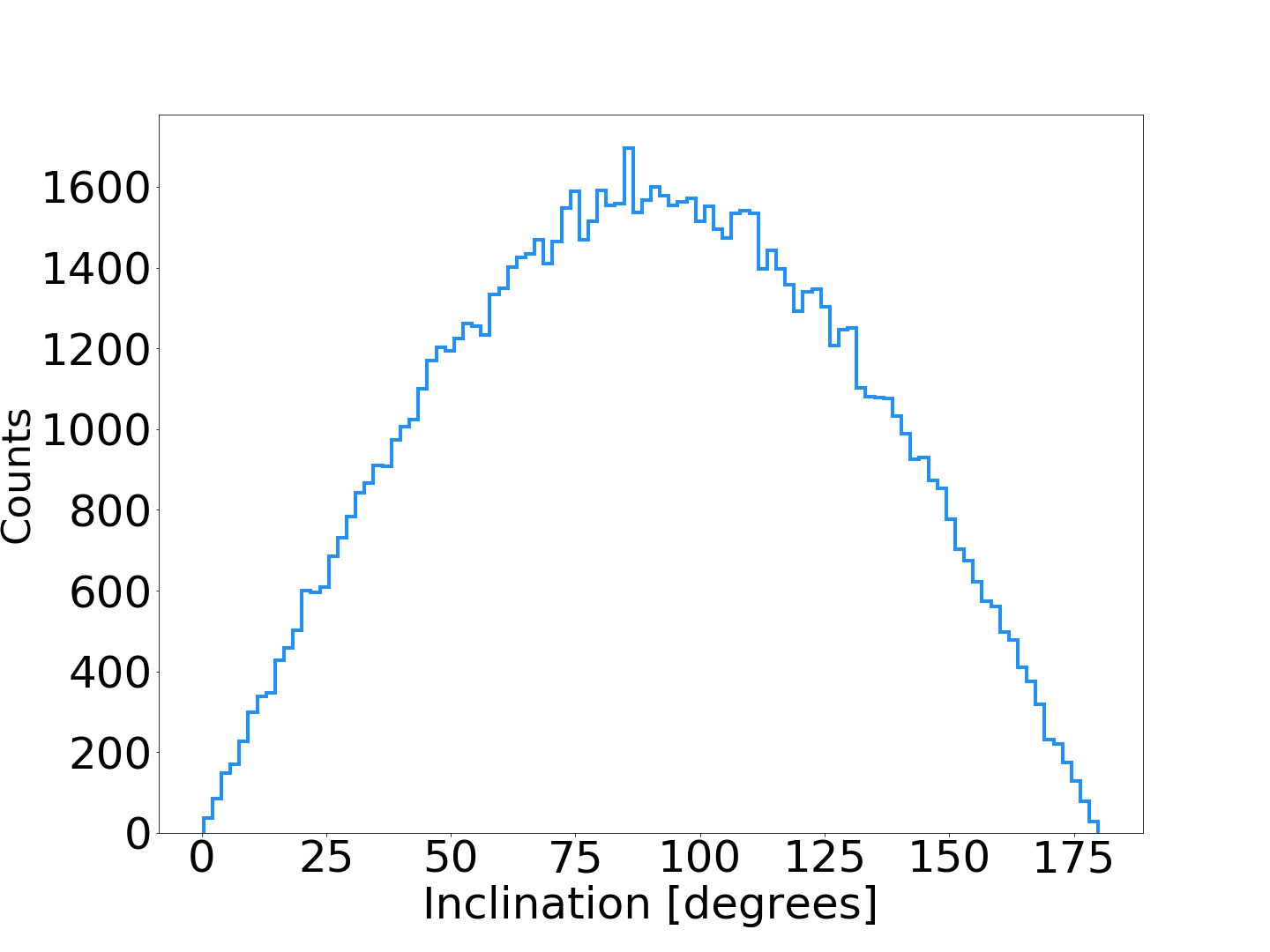}
        \end{subfigure}
        \hfill
        \begin{subfigure}[b]{0.45\textwidth}
            \centering
            \includegraphics[width=1\textwidth]{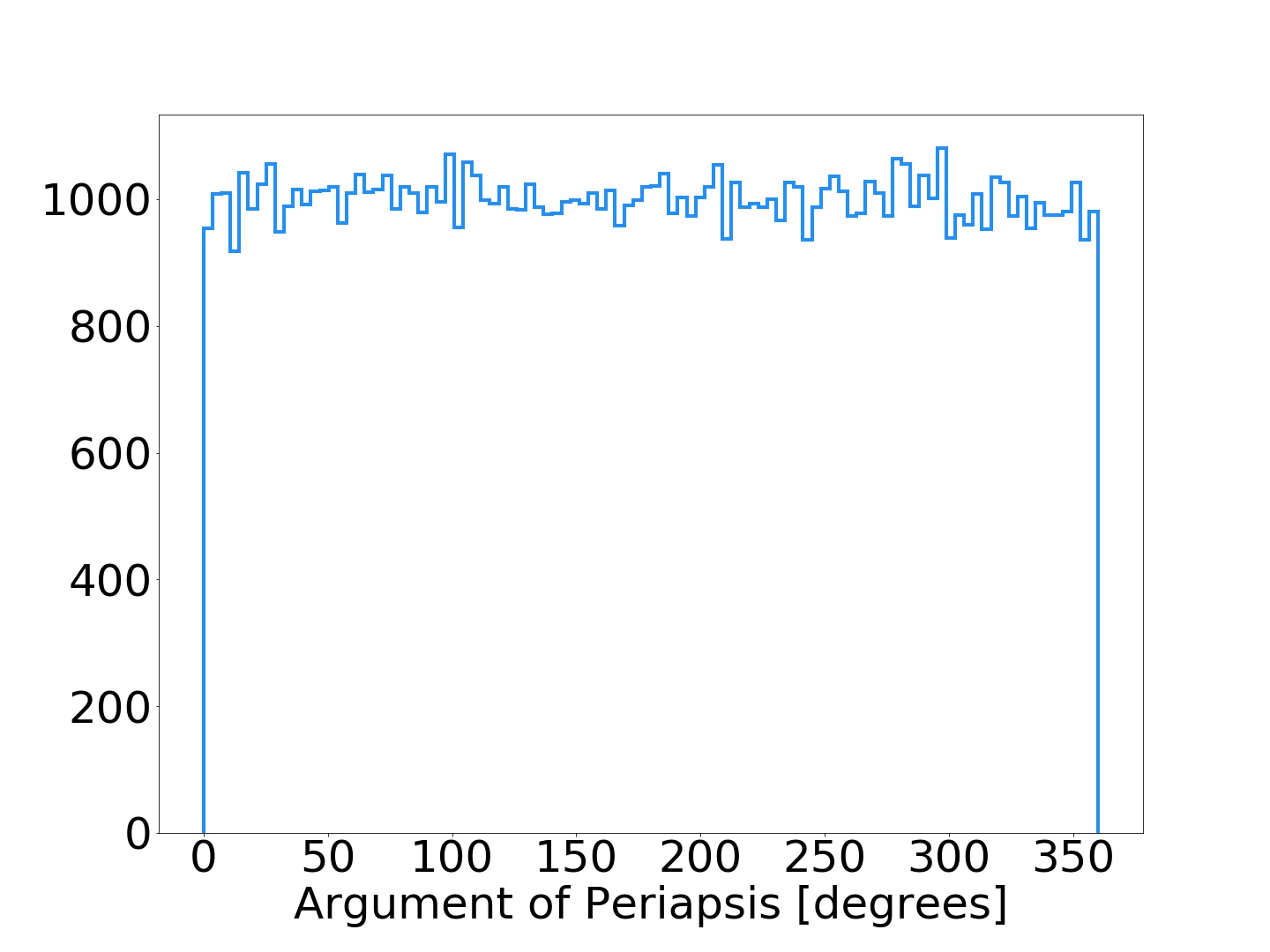}
        \end{subfigure}
        \hfill
        \begin{subfigure}[b]{0.45\textwidth}
            \centering
            \includegraphics[width=1\textwidth]{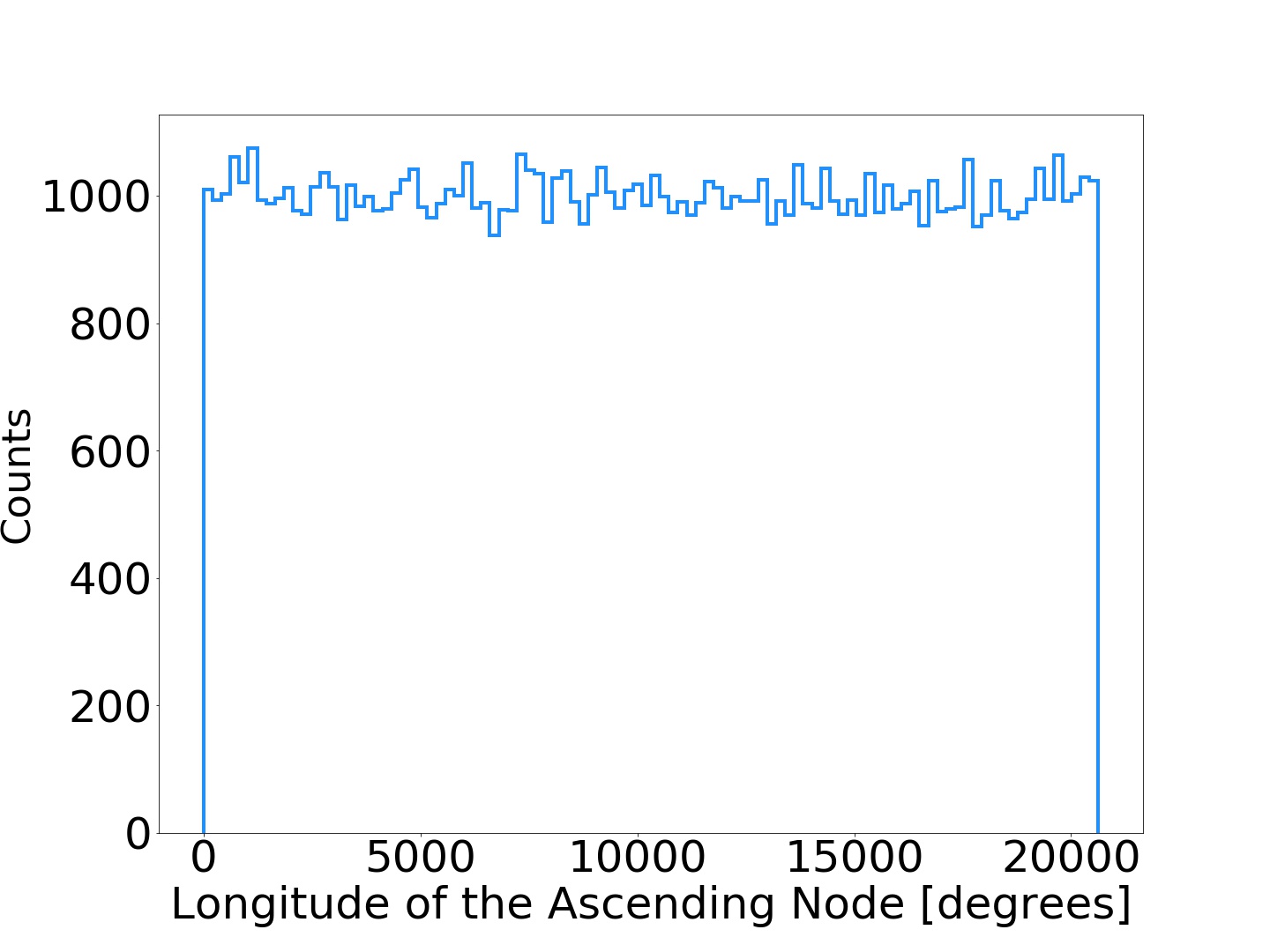}
        \end{subfigure}
        \hfill
        \begin{subfigure}[b]{0.45\textwidth}
            \centering
            \includegraphics[width=1\textwidth]{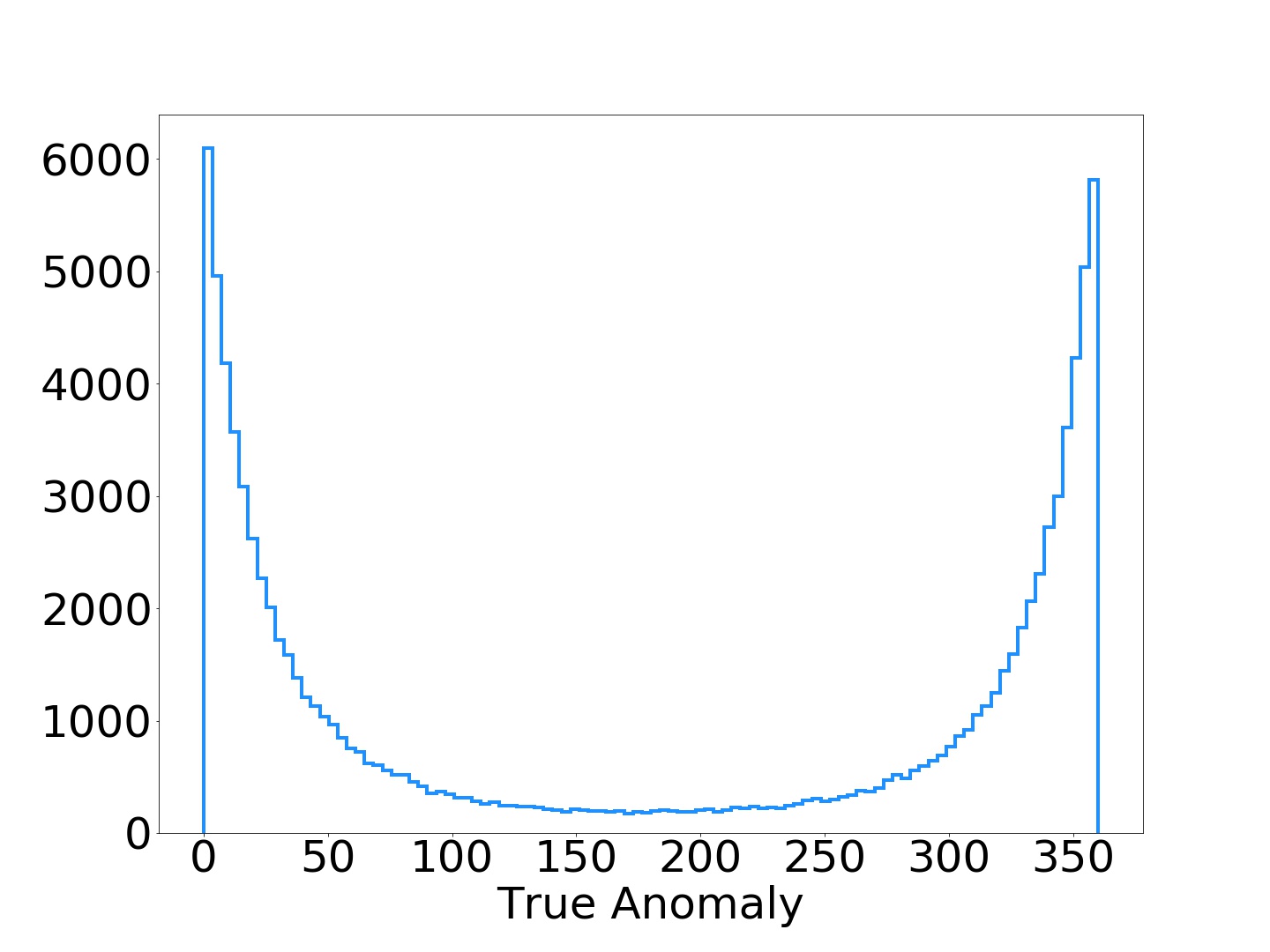}
        \end{subfigure}
          
    }
    
     \caption{Distributions of orbital elements used in Sect. \ref{subsec:isoclus} to the generate isotropic cluster, including the semimajor axis, eccentricity, inclination, argument of periapsis, longitude of the ascending node, and true anomaly. By definition, all stars in the cluster are bound to the potential of Sgr~A*. }
     \label{fig:isotropicdisthisto}
   \end{figure}

\end{appendix}

\end{document}